\documentclass[preprint2]{proto}
\usepackage{times}
\usepackage{xcolor}
\usepackage{graphicx}
\usepackage{float}
\usepackage{dblfloatfix}
\usepackage{natbib,amssymb,amsmath}
\usepackage{soul}
\usepackage{hyperref}

\bibpunct{(}{)}{;}{a}{}{,}

\newcommand*\mean[1]{\overline{#1}}

\usepackage{xcolor}

\begin{document}

\title{\textbf{\LARGE “Dust Emission and Dynamics”}}

\author {\textbf{\large Jessica Agarwal}}
\affil{\small\em Technische Universit\"at Braunschweig, Germany}
\affil{\small\em Max-Planck-Institut f\"ur Sonnensystemforschung, Germany}

\author {\textbf{\large Yoonyoung Kim}}
\affil{\small\em Technische Universit\"at Braunschweig, Germany}

\author {\textbf{\large Michael S. P. Kelley}}
\affil{\small\em University of Maryland, USA}

\author {\textbf{\large Raphael Marschall}}
\affil{\small\em Southwest Research Institute, USA}
\affil{\small\em CNRS, Observatoire de la C\^ote d'Azur, Laboratoire J.-L. Lagrange, France}

\begin{abstract}

  \begin{list}{ } {\rightmargin 1in}
    \baselineskip = 11pt
    \parindent=1pc
    {\small
      When viewed from Earth, most of what we observe of a comet is dust.
The influence of solar radiation pressure on the trajectories of dust particles depends on their cross-section to mass ratio. Hence solar radiation pressure acts like a mass spectrometer inside a cometary tail.
The appearances of cometary dust tails have long been studied to obtain information on the dust properties, such as characteristic particle size and initial velocity when entering the tail. Over the past two decades, several spacecraft missions to comets have enabled us to study the dust activity of their targets at much greater resolution than is possible with a telescope on Earth or in near-Earth space, and added detail to the results obtained by the spacecraft visiting comet 1P/Halley in 1986. We now know that the dynamics of dust in the inner cometary coma is complex and includes a significant fraction of particles that will eventually fall back to the surface. The filamented structure of the near-surface coma is thought to result from a combination of topographic focussing of the gas flow, inhomogeneous distribution of activity across the surface, and projection effects. It is possible that some larger-than-centimetre debris contains ice when lifted from the surface, which can affect its motion. Open questions remain regarding the microphysics of the process that leads to the detachment and lifting of dust from the surface, the evolution of the dust while travelling away from the nucleus, and the extent to which information on the nucleus activity can be retrieved from remote observations of the outer coma and tail.
    \\~\\~\\~}
  \end{list}
\end{abstract}

\section{Introduction}
\label{sec:intro}
Dust released from the surface is the most observationally accessible constituent of a comet. When a comet becomes visible to the naked eye, we see primarily sunlight scattered by dust particles in its coma and tail.

Theories explaining the appearance and formation of comet tails date many centuries back. Our concepts to constrain dust properties like size distribution, ejection times and velocities from the shape of and brightness distribution in a comet tail have their origins in the 19th century and primarily exploit the size dependence of solar radiation pressure on dust (Section~\ref{sec:radpr}). These concepts are described in detail in the Comets II book chapter by \cite{fulle-cometsII-2004} and mainly yield information on the properties of the dust as it leaves the sphere of influence of the nucleus.

The strongest limit of this approach is set by the finite resolution of the telescope images available. For a ground-based telescope, the typical seeing-limited resolution is of the order of 1\arcsec, which corresponds to 725\,km at a comet-observer distance of 1 au. Under exceptional circumstances, such as Hubble Space Telescope observations during a close (0.1 au) Earth flyby, a spatial resolution of a few kilometers can be achieved \citep[e.g., ][]{li-kelley2017}. However, no contemporary Earth- or near-Earth-based telescope can resolve the nucleus of a comet and the coma immediately above its surface.

The images returned by ESA's Giotto spacecraft from comet 1P/Halley in 1986 \citep{keller-arpigny1986} were the first to resolve this innermost part of the coma. They showed the comet nucleus as a solid object, and the dust as it emerges from its surface. It became clear that cometary nuclei were highly irregular bodies, certainly in shape and potentially in composition, and also that the brightness of the innermost dust coma was spatially highly variable. It displayed bright linear features apparently emanating from the surface, embedded in a more diffuse background (Section~\ref{subsec:jets}). 

Since the publication of the Comets II book, major progress in understanding the motion of dust in the near environment of cometary nuclei was enabled by several space missions. Their returned images confirmed the 1P/Halley results of the detailed fine structure in the coma brightness distribution. Indications were found that the debris itself can be outgassing and subject to physical evolution, and that part of it falls back to the surface. But open questions remain in particular concerning how activity is distributed across the surface, why it is spatially and temporally variable, and which microphysical processes lead to the ejection of dust. The answers to all these questions are necessary to understand the interior structure and composition of cometary nuclei, and eventually their formation.

ESA's Rosetta mission provided us with a comprehensive, 2-year body of data obtained with various complementary techniques, including imaging and spectroscopy of dust and the surface at various wavelengths, and in situ analysis of the composition, density and velocity of both gas and dust. These data represent the best constraints we have to understand how, where and when dust is released from a cometary surface and its subsequent journey back to the surface or to interplanetary space.

In this chapter, we first review the forces considered relevant for this outward journey of a dust particle (Section~\ref{sec:forces}). We outline the still rudimentary knowledge of how dust activity is distributed across the comet surface and what methods can help to address this question (Section~\ref{subsec:surface_emission}). We next describe how the dust particles are accelerated in the gas flow (Section~\ref{sec:acczone}), how solar gravity and radiation pressure take over as the gas dilutes (Section~\ref{sec:outercoma}) and finally address the motion of dust in the comet's tail and trail, and its transition to the zodiacal cloud (Section~\ref{sec:tailtrail}). Open questions and potential means to address them are discussed in Section~\ref{sec:future}.

\section{Forces acting on dust}
\label{sec:forces}

\subsection{Gravity and tidal forces}
\label{subsec:gravity}
The gravitational force of a massive body acting on a particle of mass $m_{\rm d}$ is
\begin{equation}
  \label{eq:grav_nucl}
F_{\rm g} = m_{\rm d} g,
\end{equation}
where $g$ is the gravitational acceleration of that body.

At distance $r$ from the nucleus center of mass, the gravitational acceleration by the nucleus mass, $M$, is defined as
\begin{equation}
  \label{eq:grav_acc}
  g = \frac{G M}{r^2},
\end{equation}
where $G$ is the gravitational constant.
In the immediate environment of the nucleus, the gravitational field cannot be approximated by that of a point mass as in Eq.~\ref{eq:grav_nucl}. The spatial extent of the nucleus, in combination with its irregular shape and potentially inhomogeneous internal mass distribution, will necessitate to consider also higher orders of the gravitational potential for calculations of the dust motion \citep[e.g., ][]{werner1994}. 

The solar gravitational force on a particle at heliocentric distance $r_{\rm h}$ is
\begin{equation}
  \label{eq:grav_sun}
  F_{\rm g} = \frac{G M_\odot}{r_{\rm h}^2} m_{\rm d},
\end{equation}

where $M_\odot$ is the mass of the Sun.
The sphere of gravitational influence (Hill sphere) of the comet nucleus is estimated as
\begin{equation}
  \label{eq:Hill_radius}
  R_{\rm {Hill}} = r_{\rm h} \left( \frac{1}{3} \frac{M}{M_\odot} \right)^{1/3}.
\end{equation}
Thus, for a typical 2-km radius Jupiter-family comet with a bulk density $\rho_{\rm n}$ = 500 kg m$^{-3}$, perihelion distance of 1 au, and aphelion at 6 au, the Hill radius will range from roughly 200 km at perihelion to 1300 km at aphelion.

The radial velocity component required for an object at distance $r$ from mass $M$ to stop being gravitationally bound to this mass (escape speed) is given by
\begin{equation}
  \label{eq:v_esc}
v_{\rm esc} = \sqrt{\frac{2 G M}{r}}.
\end{equation}
Typical escapes speeds from the surfaces of km-sized bodies are of order 1 m s$^{-1}$.

For the motion of a dust particle in a frame attached to the nucleus center of mass, the difference in solar gravity between the locations of the particle and of the nucleus (the tidal force) is more relevant than the absolute value of solar gravity. A particle of mass $m_{\rm d}$ located on the Sun-nucleus line and at a distance $r$ from the nucleus in either direction is subject to a tidal force directed away from the nucleus given by
    \begin{equation}
      F_{\rm tr} = \frac{2 G M_\odot}{r_{\rm h}^3} m_{\rm d}  r,
      \label{eq:tides_radial}
    \end{equation}
    while the tidal force on a particle located above the terminator points towards the nucleus with the magnitude
    \begin{equation}
      F_{\rm tl} = \frac{G M_\odot}{r_{\rm h}^3} m_{\rm d} r.
      \label{eq:tides_lateral}
    \end{equation}

\subsection{Drag by surrounding gas coma}
\begin{figure*}
  \includegraphics[width=\textwidth]{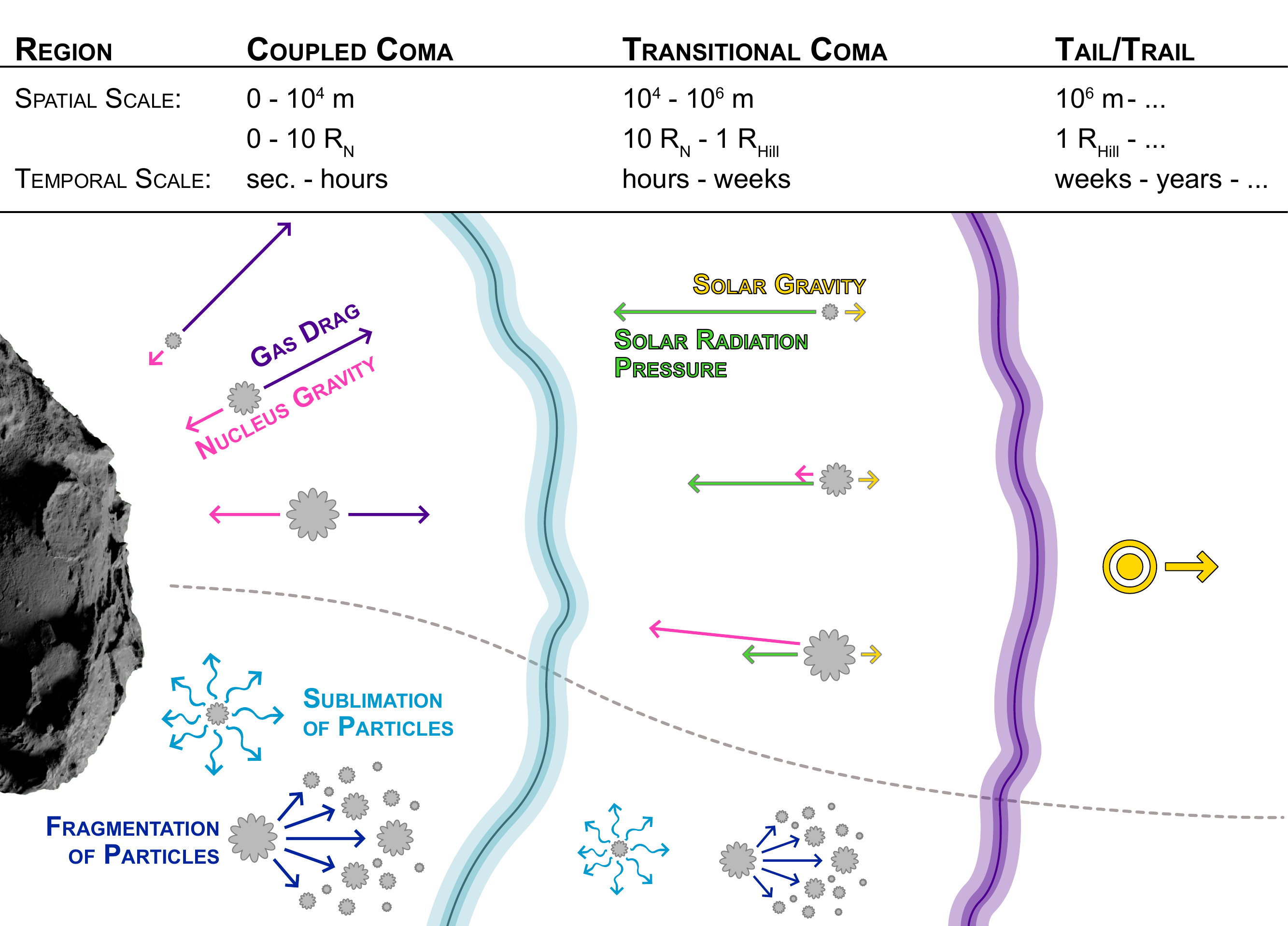}
  \caption{Sketch of the cometary dust environment which we divide into three main dynamical regions: 1) The coupled coma region where the dust dynamics is dominated by local forces connected to the nucleus (gas drag and nucleus gravity) (Section~\ref{sec:acczone}); 2) the transitional coma region where the dust has decoupled from the gas and small particles transition to being dominated by solar forces (gravity and radiation pressure) while large particles are bound in the gravitational field of the nucleus (Section~\ref{sec:outercoma}); 3) the dust tail and trail within which the escaping dust particles are purely governed by solar forces (Section~\ref{sec:tailtrail}). The boundaries between these regions are complex 3D surfaces. The spatial and temporal scales given are rough estimates for a 67P-like comet at 1 au. $R_{\rm n}$ is the nucleus radius, and $R_{\rm Hill}$ the Hill radius. This figure is a reproduction of Fig. 4 in \cite{marschall-skorov2020}.}
  \label{fig:dustForcesAndProcesses}
\end{figure*}

The main force accelerating dust particles away from the nucleus surface comes from their interaction with the surrounding gas field.
As illustrated in Fig.~\ref{fig:dustForcesAndProcesses}, the drag force is strongest within $\sim$10 nucleus radii, $R_{\rm n}$.
By that distance the gas densities have diluted significantly, making molecule-dust collisions rare and momentum transfer from the gas flow to the dust particles inefficient. The gas dynamics are discussed by Marschall et al. in this book.

Once a dust particle is detached and lofted from the surface, the surrounding gas molecules collide with it and accelerate the particle.
The main direction of gas expansion is away from the nucleus surface and therefore so is the net force on the dust particles.

When dust densities are low enough that particles do not exert a back reaction onto the gas flow (e.g. deceleration and/or heating of the gas flow) they can be considered as test particles within the gas flow and thus be treated mathematically separately \citep{tenishev-combi2011}.
This condition is given in most cases but there are exceptions.
For example in the event of a strong outburst, where the dust plume is optically thick, this condition is certainly not satisfied.
For such cases, a multi-phase simulation needs to be adopted \cite[e.g.,][]{Shou2017} which also takes into account inter-particle collisions.

In the following, we assume the more common case where dust particles do not significantly influence the gas flow.
In addition, in most cases, the mean free path of the molecules is much larger than the dust particle size, and therefore free molecular aerodynamics can be applied.
In this scenario \citep{FinsonProbstein1968,Gombosi1985,Gombosi1986,Gombosi1987,Sengers2014,Marschall2016} the drag force, $F_{\rm D}$, on a spherical particle is
\begin{equation} \label{eq:drag-force}
  \vec{F}_{\rm D} = \frac{1}{2} C_{\rm D} m_{\rm g} n_{\rm g} \sigma_{\rm d} \left| \vec v_{\rm g} - \vec v_{\rm d} \right| \left( \vec v_{\rm g} - \vec v_{\rm d} \right),
\end{equation}
where $\sigma_{\rm d}$ is the geometric cross-section of the dust particle, and $\vec{v}_{\rm d}$ is its velocity.
The mass of a gas molecule is $m_{\rm g}$, and $n_{\rm g}$ and $\vec v_{\rm g}$ are their number density and macroscopic velocity, respectively.
$C_{\rm D}$ is called the drag coefficient.
For an equilibrium gas flow, and a mean free path of the molecules much larger than the dust size, the drag coefficient \citep{Bird94} is defined as
\begin{equation} \label{eq:drag-coefficient}
  \begin{split}
    C_{\rm D} = &\frac{2 \zeta^2 + 1}{\sqrt{\pi} \zeta^3} e^{-\zeta^2} + \frac{4 \zeta^4 + 4 \zeta^2 - 1}{2 \zeta^4} \text{erf}(\zeta)\\
    &+ \frac{2 \left( 1 - \varepsilon \right) \sqrt{\pi}}{3 \zeta} \sqrt{\frac{T_{\rm d}}{T_{\rm g}}} ,
  \end{split}
\end{equation}
with the gas temperature $T_{\rm g}$, the dust particle temperature $T_{\rm d}$, the fraction of specular reflection $\varepsilon$, and the molecular speed ratio
\begin{equation} \label{eq:zeta}
  \zeta = \frac{ \left| \vec v_{\rm g} - \vec v_{\rm d} \right| }{\sqrt{\frac{2k_{\rm b} T_{\rm g}}{m_{\rm g}}}},
\end{equation}
where $k_{\rm b}$ is the Boltzmann constant. 

\begin{figure}
  \includegraphics[width=0.5\textwidth]{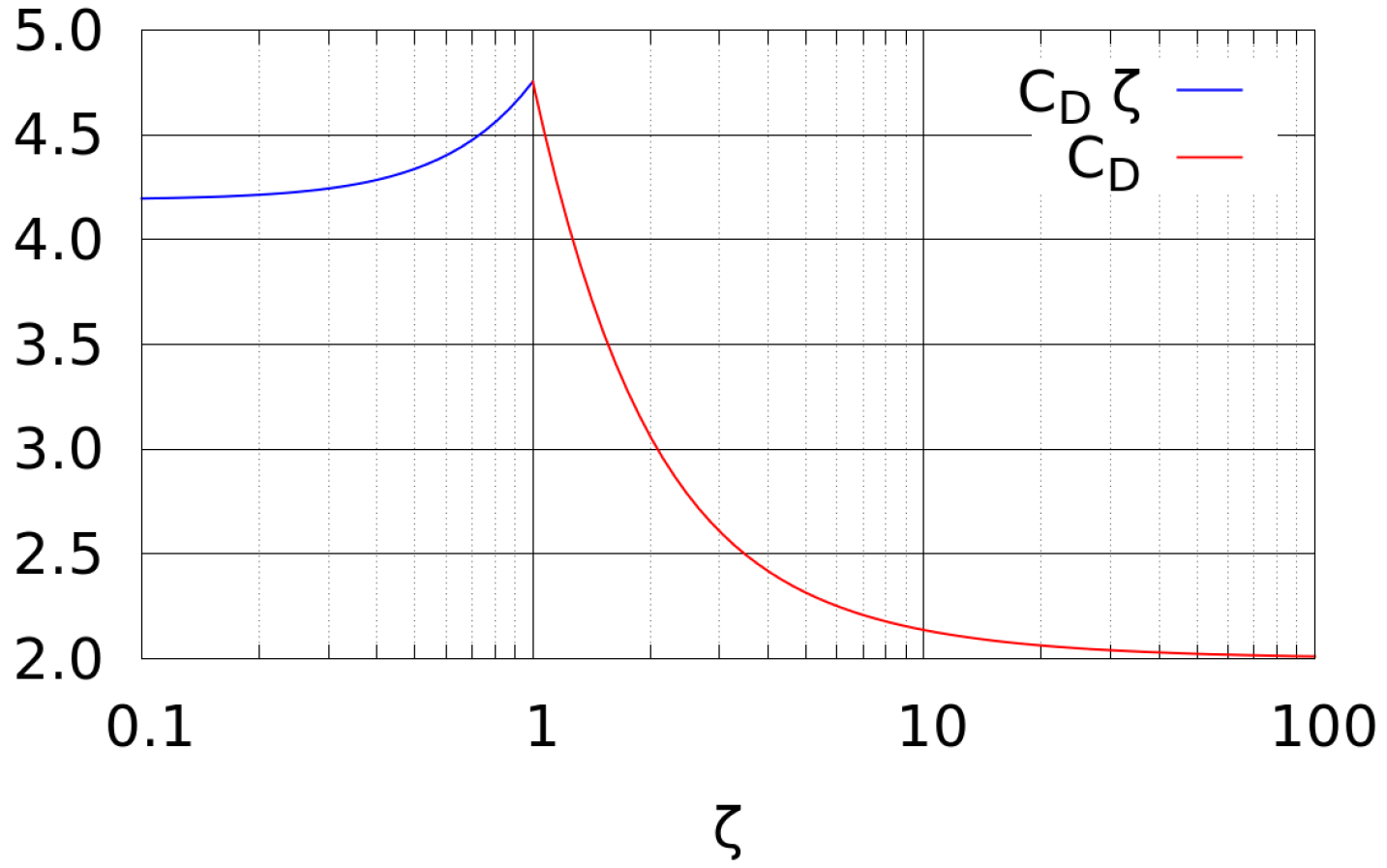}
  \caption{Drag coefficient $C_{\rm D}$ as a function of the parameter $\zeta$ as given by Eqs.~\ref{eq:drag-coefficient} and \ref{eq:zeta} and assuming $T_{\rm d}=T_{\rm g}$ and $\varepsilon$=0.}
  \label{fig:dragCoeff}
\end{figure}

Figure~\ref{fig:dragCoeff} shows the drag coefficient as a function of $\zeta$.
For large particles the dust speed is much smaller than the gas speed. At the surface, the gas speed is of the same order as the thermal speed (the denominator in Eq.~\ref{eq:zeta}) and therefore $\zeta \sim 1$ and $C_{\rm D} \sim 5$ (Fig.~\ref{fig:dragCoeff}). 
On the other hand, for a typical comet (water dominated emission well within the snow line) at large cometocentric distances (e.g., at $10R_{\rm n}$), the temperature of the gas has cooled to a few tens of Kelvin but the gas speed has reached the order of 1 km/s. In this case slowly moving particles have $\zeta \sim 10$ and thus $C_{\rm D} \sim 2$ (Fig.~\ref{fig:dragCoeff}).
Very small, typically sub-micron, particles, may attain a significant fraction of the gas speed already very close the surface, such that $\zeta < 1$  and $C_{\rm D} > 5$ (Fig.~\ref{fig:dragCoeff}).
Because the drag coefficient asymptotically approaches 2 for most particle sizes, a size-independent $C_{\rm D} =2$ is often assumed rather than the more complicated Eq.~\ref{eq:zeta}. This choice implies that the acceleration of in particular small particles is underestimated.

Eq.~\ref{eq:drag-coefficient} describes the idealized case of spherical particles in a gas flow when the mean free path of the gas molecules is larger than the dust particle size.
Real dust particles are not spherical and, in particular, larger dust grains are porous and fluffy aggregates \citep{Kolokolova10b,Schulz2015,rotundi15,Langevin2016,Bentley2016,Mannel2016,Levasseur2018}.
This affects the dynamics of the particles as shown by \cite{Skorov2016,Skorov2018}.
They found that porous aggregates are accelerated to significantly higher speeds than their compact counterparts of the same radius.
This behavior can be mimicked in the spherical particle paradigm described above by adjusting the masses and geometric cross-sections to the respective values of the porous particles.
In this sense, the spherical particles can be understood as effective particles with the given mass and cross-section.

Additionally, particles have been observed to rotate in the comae of comets 103P and 67P \citep{hermalyn13, fulle-ivanovski2015}, and the effect of oblate or prolate particle shapes on their dynamics was studied by \cite{Ivanovski2017a,Ivanovski2017b}.
Not only will non-spherical particles begin to rotate in the gas flow but they may also accelerate to higher speeds than spherical ones.
Unless a particle rotates sufficiently fast, the influence of particle rotation on their dynamics cannot simply be parameterized into a spherical particle paradigm (Eq.~\ref{eq:drag-force}) as described above for porous particles, because spherical particles only experience a force along the direction of the gas flow, while non-spherical, rotating particles also experience a force perpendicular to the flow.
This component of the acceleration is not included in Eq.~\eqref{eq:drag-force}.

\subsection{Intrinsic outgassing}
\label{sec:outgassing}

Solid particles in the coma can contain volatile ices.  When the grain temperature is sufficiently warm, the ices will sublimate.  This loss of vapour  accelerates the particle.  In the ideal case of an isotropically outgassing spherical particle, the net acceleration is 0.  However, coma particles are not spheres, and outgassing may not be uniform, especially from large particles that can sustain a temperature gradient across their surfaces.  Rapid rotation of a particle does help distribute the absorbed sunlight across the surface, and smooths out the temperature gradient, but the spin axis may prevent illumination of some portions of the grain.  As a result, the acceleration will typically have an anti-solar component.  Whether or not a tangential component exists, depends on the spin state, shape, porosity, and thermal characteristics of the materials.

The calculation of the acceleration from outgassing is analogous to the non-gra\-vi\-ta\-tion\-al acceleration of cometary nuclei \citep{marsden73}, except that the microphysics of the particle, e.g., thermal and radiative properties, are very different.  Setting these complexities aside,
the force from gas sublimation (also called ``rocket force''), $F_{\rm s}$ can be calculated from:
\begin{equation}
  F_{\rm s} = \sigma_{\rm d} m_{\rm g} Z v_{\rm th} \kappa f_{\mathrm{ice}} ,
  \label{eq:rocket}
\end{equation}
where $Z$ is the number sublimation rate of the ice per surface area, $\kappa$ is the degree of asymmetry of outgassing (0 for isotropic, 1 for directly toward the Sun), $f_{\mathrm{ice}}$ is the effective fractional surface area of the ice \citep[cf.][]{kelley13}, and
\begin{equation}
v_{\rm th} = \frac{\pi}{4} \sqrt{\frac{8 k_{\rm b} T_{\rm g}}{\pi m_{\rm g}}}
\end{equation}
is the mean thermal expansion speed of the gas.

The details hidden within $\kappa$ may be complex, but Eq.~\ref{eq:rocket} is still useful for estimating the potential order of magnitude of acceleration by outgassing.

Since the force is generally opposite the Sun, it works similarly to acceleration by radiation pressure, in that the particles are accelerated into the tail direction.  For example, \citet{reach09} examined the acceleration from outgassing water ice at 1~au and found that the resultant force, $F_{\rm s}$, can be comparable to or much stronger than that from solar radiation $F_{\rm r}$ (Section~\ref{sec:radpr}), and estimated $F_{\rm s}/F_{\rm r}\sim100$ for a 1-mm sized grain with $f_{\mathrm{ice}}=0.0017$.  Over small heliocentric distance ranges and when the relevant ice is undergoing free sublimation, the force may be approximated as equivalent to the $\beta$-parameter, i.e., a force $\propto r_{\rm h}^{-2}$ that is addressed with a reduced-gravity solution (Section~\ref{sec:radpr}).

Likely, the outgassing force is observable only for centimetre-sized and larger particles, because smaller particles lose their ice content too quickly and hence too close to the surface for the non-gravitational effect on their trajectories to be measured. This conclusion was reached by \cite{markkanen-agarwal2020} and \cite{davidsson-birch2021} from thermophysical modelling, and, complementary, by \cite{reach09} from studying the motion of particles in the debris trail of comet 73P/Schwassmann-Wachmann 3.

It is possible that asymmetric outgassing changes the rotation rate of an ice-bearing comet fragment, which can lead to disintegration by centrifugal force on time scales of less than a day for a meter-sized fragment \citep{jewitt_332P-1}.
\cite{steckloff-jacobson2016} propose that fragment disintegration by sublimation torques can lead to the formation of striae observed in the tails of some enigmatic, bright comets. Alternative explanations of striae formation invoke electromagnetic forces (Section~\ref{sec:em_forces}).

\subsection{Solar radiation pressure}
\label{sec:radpr}
The radiation force is proportional to the solar intensity multiplied by the cross-sectional area of the particle:
\begin{equation}
  \label{eq:radpr}
  F_{\rm r} = \frac{Q_{\rm pr}}{c} \left( \frac{L_\odot}{4 \pi r_{\rm h}^2} \right) \sigma_{\rm d},
\end{equation}
where $L_\odot$ is the solar luminosity, $c$ is the speed of light, and $Q_{\rm pr}$ is the radiation pressure coefficient averaged over the solar spectrum \citep{burns-lamy1979}.
Dynamical models commonly use a simplified treatment of comet dust, assuming $Q_{\rm pr}$ = 1. A more detailed treatment dependent on dust mineralogies and structures (compact or fluffy) is discussed in Kolokolova et al. in this volume.

Since the solar radiation force opposes the solar gravity and both forces are proportional to $1/{r_{\rm h}^2}$, the net force can be considered as reduced solar gravity,

\begin{equation}
  F_{\rm net} = F_{\rm r} - F_{\rm g} = (1-\beta) F_{\rm g} ,
\end{equation}
where the $\beta$ parameter is the ratio of the radiation force, $F_{\rm r}$, to the solar gravitational force, $F_{\rm g}$,
\begin{equation}
  \label{eq:beta}
  \beta \equiv \frac{F_{\rm r}}{F_{\rm g}} = \frac{3 L_{\odot}Q_{\rm pr}}{16\pi G M_\odot c \rho_{\rm d} a} \quad \!\!  = C_\beta \frac{Q_{\rm pr}} {\rho_{\rm d} a},
\end{equation}
with $C_\beta = 5.77\!\times\!10^{-4} \,\mathrm{kg \,m^{-2}}$, and $a$ representing the radius of a spherical particle. More generally, $1/a$ can be expressed as $4\,\sigma_{\rm d} \rho_{\rm d}/ (3 m_{\rm d})$, with $\rho_{\rm d}$ being the dust bulk density.

Calculations show that silicate particles tend to have $\beta<$1 regardless of aggregate structure \citep{Silsbee2016}, while absorbing particles may have $\beta >$1 \citep{Kimura2016}.

\subsection{Poynting-Robertson drag}
Small particles in orbit about the Sun are influenced also by radiation pressure tangential to their motion \citep{Robertson1937,Wyatt1950}. The resulting Poynting-Robertson force is given by
\begin{equation}
  \label{eq:PR}
  F_{\rm PR} = \frac{a^2 L_{\odot}}{4c^2} \sqrt{\frac{G M_\odot}{r_{\rm h}^5}}.
\end{equation}

The Poynting-Robertson effect causes mm-sized dust particles in the zodiacal cloud (Sec.~\ref{subsec:zodiacal}) to spiral into the Sun on timescales $\tau_{\rm PR} \gtrsim 6\times10^{5}$~yr \citep{kasuga-jewitt2019}.

\subsection{Electromagnetic forces}
\label{sec:em_forces}
A dust particle embedded in the cometary or solar wind plasma and interacting with solar ultraviolet radiation is subject to charging by electron and ion collection, and secondary electron and photoelectron emission. Over time, the dust particle will assume the potential at which the involved currents balance. This equilibrium potential depends on the properties of the plasma environment and the dust particle, such as composition and surface roughness \citep{horanyi1996}. In interplanetary space, the dominant charging process is photoelectron emission, and typical dust potentials, $U$, range between 0.5V and 14V \citep{mukai1981}. A canonical value of $U$=5V is often used \citep[e.g., ][]{sterken-altobelli2012, kramer-fernandez2014}.
Solar wind interaction with the plasma tail is discussed in G\"otz et al. in this volume.

For a given surface potential and volume, the shape dependence of a grain's charge can be described by the dimensionless parameter $\kappa_{\rm e} >$1 that is minimal for a sphere ($\kappa_{\rm e}$=1) and can reach values up to $\kappa_{\rm e}$=5 for fractal particles \citep{auer-kempf2007}. The integrated charge of a grain can thus be described as
\begin{equation}
  \label{eq:charge}
  q = 4 \pi \varepsilon_0 a \kappa_{\rm e} U,
\end{equation}
where $a$ is the radius of a volume-equivalent sphere, and $\varepsilon_0$ is the electric permittivity in vacuum.

In the presence of a magnetic field, $\vec{B}$, a charged particle moving with velocity $\vec{v}$ relative to the field is subject to the Lorentz force
\begin{equation}
  \vec{F}_{\rm L} = q (\vec{v} \times \vec{B}).
  \label{eq:lorentz1}
\end{equation}
Outside the immediate environment of the comet, the relevant field is the interplanetary magnetic field (IMF), and the velocity of the dust particle relative to this field can be approximated by the velocity of the solar wind ($v_{\rm SW}$=400-800 km/s radially outward from the Sun) that carries the magnetic field and is at least an order of magnitude faster than typical heliocentric velocities of comets. Splitting the IMF into a radial ($B_r$), an azimuthal ($B_\phi$) and a normal component ($B_\theta$=0), Eq.~\ref{eq:lorentz1} reduces to \citep{kramer-fernandez2014}
\begin{equation}
  F_{\rm L} = \pm q v_{\rm SW} B_\phi = q v_{\rm SW}  B_{\phi,0} \frac{r_0}{r_{\rm h}} \cos{\beta_{\rm hg}},
  \label{eq:lorentz2}
\end{equation}
where $\beta_{\rm hg}$ is the heliographic latitude, $r_0$=1 au, and $B_{\phi,0}$= 3 nT is the azimuthal field strength at 1 au. Hence, the Lorentz force decreases with heliocentric distance as 1/$r_{\rm h}$, less steeply than solar gravity and radiation pressure. For a given type of particle, the relative importance of the Lorentz force will increase with heliocentric distance (Fig.~\ref{fig:force_vs_rh}). \cite{kramer-fernandez2014} and \cite{hui-farnocchia2019} reported that including the Lorentz force in a model of the dust motion significantly improved reproducing the orientation of the dust tails of comets Hale-Bopp and C/2010 U3 (Boattini) at heliocentric distances between 15 and 30 au.

\cite{price-jones2019} find that changes in the appearance of striae in the tail of comet C/2006 P1 (McNaught) coincided with the comet crossing the heliospheric current sheet and infer that dust in the striae was charged and hence subject to the Lorentz force. Striae are linear features inside a comet's dust tail of unknown origin. They are only seen in comets with very high production rate, typically dynamically new comets. Alternative models of striae formation favour  processes of instantaneous disintegration of, e.g., highly non-spherical grains \citep{sekanina-farrell1980} or of large boulders fragmenting under outgassing-induced torques \citep[][see also Section~\ref{sec:outgassing}]{steckloff-jacobson2016}.

\cite{Fulle2015} find that near the Rosetta spacecraft, fluffy dust particles of extremely low density may get charged by secondary electrons from the spacecraft and disintegrate, leading to the detection of particle swarms by the on-board dust instrument GIADA. In the vicinity of comet 67P, also charged water clusters smaller than 100~nm were detected \citep{gombosi-burch2015}.

\subsection{Electrostatic lofting}
          Charged dust lofting and transport have been proposed to explain observations of airless bodies in the solar system, such as the lunar horizon glow \citep{Rennilson1974}, the ``spokes'' in Saturn's rings \citep{Morfill1983}, and dust ponds formed on asteroid Eros \citep{Colwell2005}.
There is little published work on the relevance of this effect in the presence of outgassing (such as on an active comet), but \cite{Nordheim2015} modeled the electrostatic charging of the nucleus of comet 67P and showed that charged dust grains with radii $<$50~nm may be electrostatically ejected from the nucleus in situations of weak activity.
  The electrostatic force on a particle is
  \begin{equation}
    \label{eq:electrostatic}
    F_{\rm ES} = qE,
  \end{equation}
  where $E$ is the local electric field strength. Details on the particle charging equations are presented in \citet{Zimmerman2016} and \citet{Wang2016}.
  
We here summarize some aspects of dust charging on non-cometary airless bodies:
  Dust on the lunar surface is levitated due to electrostatic charge gradients resulting from uneven solar illumination. Because on km-sized asteroids gravity is much lower, dust can be electrostatically ejected from such bodies \citep{Lee1996}.

Recent asteroid missions have observed rocky surfaces on asteroids Bennu and Ryugu, indicating a lack of regolith \citep{Jaumann2019, Lauretta2019}. These observations and the particle ejection events observed on Bennu may be partly caused by electrostatic dust lofting and escape \citep{Hartzell2022,nichols-scheeres2022}.

\subsection{Inter-particle cohesion}
\label{subsec:cohesion}
Inter-molecular (e.g., Van der Waals) forces between the surfaces of neighbouring particles or grains are held responsible for the internal strength of dust aggregates, agglomerates, chunks and surfaces. The precise form and magnitude of these forces depends on the structure and composition of the material, which are not well known.

For a lunar-type regolith surface with average grain size $a$, \cite{sanchez-scheeres2014} derive a strength of
\begin{equation}
  \label{eq:strength_regolith}
  T_{\rm reg} = C_{\rm reg} C_{\#} \phi / a,
\end{equation}
where $C_{\rm reg} = 4.5\!\times\!10^{-3}\, \mathrm{N m^{-1}}$. $C_{\#}$ is the number of neighbouring particles that a given grain touches (the coordination number), and $\phi$ is the volume filling factor of the dust layer, i.e. the fraction of the volume that is filled by matter.

For a model surface composed of agglomerates of aggregates of dust grains, and using empirical relationships based on laboratory measurements, \cite{skorov-blum2012} deduce the following expression for the tensile strength of a dust surface:
\begin{equation}
  \label{eq:strength_aggregates}
  T_{\rm agg}= C_{\rm agg} \: \phi \: \left( \frac{a}{a_0}\right)^{-2/3},
\end{equation}
with $C_{\rm agg}$ = 1.6\,Pa, and $a_0$ = 1mm.

Eqs.~\ref{eq:strength_regolith} and \ref{eq:strength_aggregates} render values that differ by two orders of magnitude at $a$=100~\micron, which illustrates the sensitive dependence on model assumptions and the lack of well constrained parameters.

\subsection{Relative importance of forces}

\begin{deluxetable}{llr}
  \tablecaption{Parameters and values used to generate Figures~\ref{fig:force_vs_size} -- \ref{fig:force_vs_rh}.}
  \tablewidth{0pt}
  \tablehead{ Quantity & Symbol & Value}
  \startdata
  Nucleus radius   & $R_{\rm n}$    & 2 km\\
  Nucleus density  & $\rho_{\rm n}$ & 500 kg m$^{-3}$\\
  Dust bulk density     & $\rho_{\rm d}$   & 500 kg m$^{-3}$\\
  Drag coefficient & $C_{\rm D}$    &  2\\
  Global water production rate\tablenotemark{a} & $Q_{\rm H_2O} (r_{\rm h}<4 {\rm au})$ & 3$\times$10$^{28}$ molecules s$^{-1}$ $(r_{\rm h} / 1 {\rm au})^{-4.5} $ \\
  & $Q_{\rm H_2O} (r_{\rm h}>4 {\rm au})$ & 4$\times$10$^{34}$ molecules s$^{-1}$ $(r_{\rm h} / 1 {\rm au})^{-15} $ \\
  Global CO$_2$ production rate\tablenotemark{b} & $Q_{\rm CO_2} (r_{\rm h})$ & 4$\times$10$^{25}$ molecules s$^{-1}$ $(r_{\rm h} / 1 {\rm au})^{-2} $\\
  Gas speed in coma       & $v_{\rm g}$    & 700 m s$^{-1}$\\
  Gas speed (from icy chunks)       & $v_{\rm th}$    & 500 m s$^{-1}$\\
  Dust speed as a function of radius, $a$,\tablenotemark{c}
  & $v_{\rm d}(a)$ & 300 m s$^{-1}$ $\sqrt{a/1 \mu {\mathrm m}}$\\
  Ice fraction in dust & $f_{\rm ice}$ & 0.01 \\
  Radiation pressure coefficient & $Q_{\rm pr}$ & 1\\
  Dust potential & $U$ & 5V\\
  Solar wind speed & $v_{\rm SW}$ & 600 km s$^{-1}$\\
  Azimuthal IMF strength at 1 au & $B_{\phi,0}$ & 3 nT\\
  Heliographic latitude & $\beta_{\rm hg}$ & 0$^\circ$\\
  Volume filling factor & $\phi$ & 0.5\\
  \enddata

  \tablenotetext{a}{The helioncentric distance dependence of the water production rate is discussed in Section~\ref{sec:turn-back}. Beyond 4 au we adopt an exponent of -15, which reasonably describes the water production rates between 5 au and 6 au if calculated from the balance of input solar energy, thermal radiation and sublimation cooling. At larger heliocentric distances, the water production rate drops even more steeply, but anyway becomes negligible compared to more volatile species.}
  \tablenotetext{b}{The CO$_2$ production rate is here assumed to be proportional to the available solar energy which scales with $r_{\rm h}^{-2}$ if radiation cooling is neglected. The true behavior of gas production rates is far more complex \citep[e.g., ][]{combi-shou2020} and variable between comets.}
  \tablenotetext{c}{The size dependence of the dust speed is analogous to Eq.~\ref{eq:vej_analytical} and Fig.~\ref{fig:dustSpeeds}.}
  
  \label{tab:dust_parameters}
\end{deluxetable}

\begin{figure*}
  \includegraphics[width=0.49\textwidth]{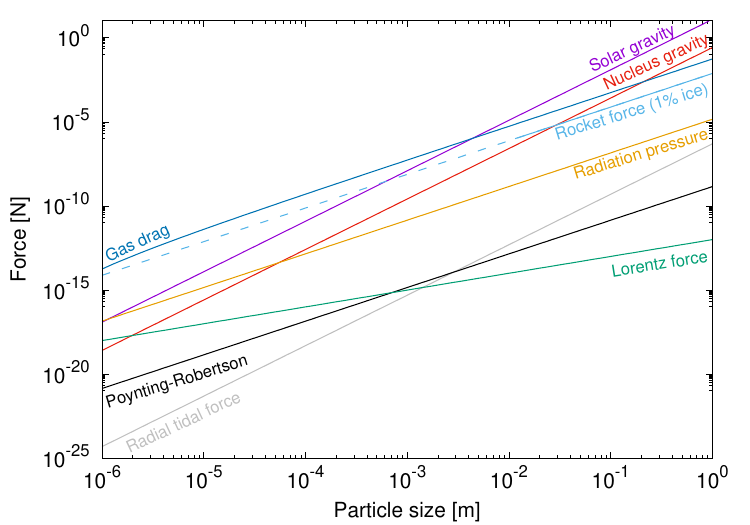}
  \includegraphics[width=0.49\textwidth]{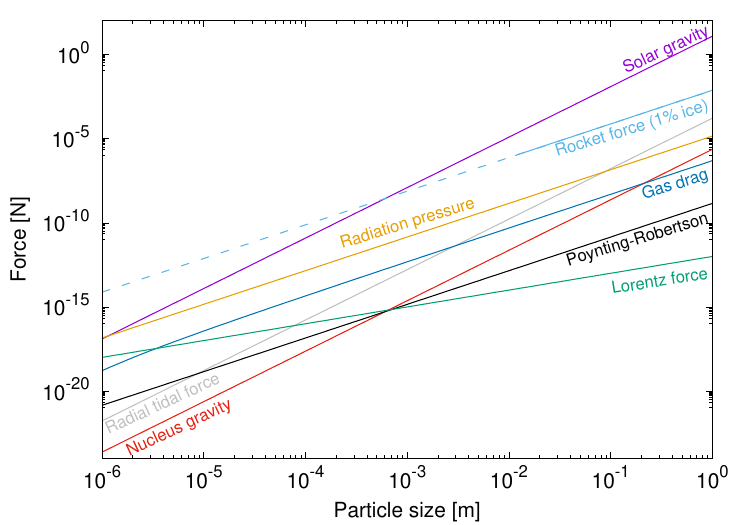}
  \caption{Forces acting on dust particles (Equations \ref{eq:grav_nucl} -- \ref{eq:grav_sun}, \ref{eq:tides_radial}, \ref{eq:drag-force}, \ref{eq:rocket}, \ref{eq:radpr}, \ref{eq:PR} -- \ref{eq:lorentz2}) as functions of particle size. The parameter values used are listed in Table~\ref{tab:dust_parameters}. The heliocentric distance is 1 au in both plots. The left panel is for a surface distance of 1\,km and displays the forces well inside the coma. The right panel refers to a surface distance of 998 km, where dust has essentially decoupled from the nucleus gravity and gas drag. The rocket force is depicted by dashed lines for sizes $<$1\,cm to indicate that ice lifetimes in small particles are too short to influence the dust dynamics on observable timescales.}
  \label{fig:force_vs_size}
\end{figure*}

\begin{figure*}
  \includegraphics[width=0.49\textwidth]{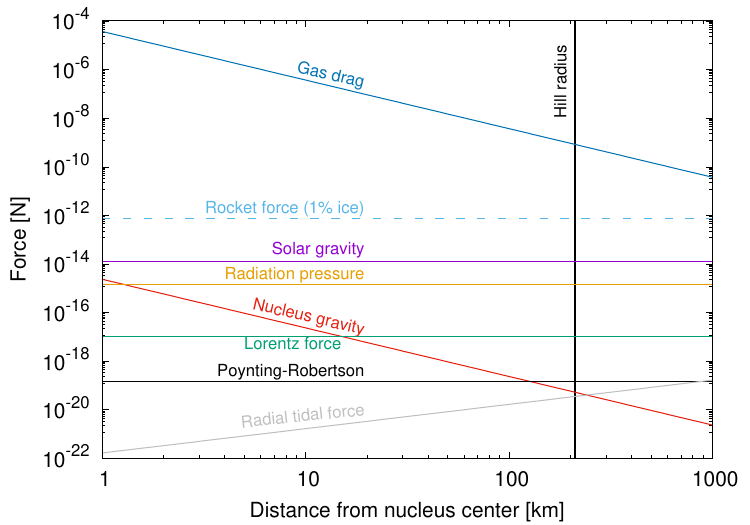}
  \includegraphics[width=0.49\textwidth]{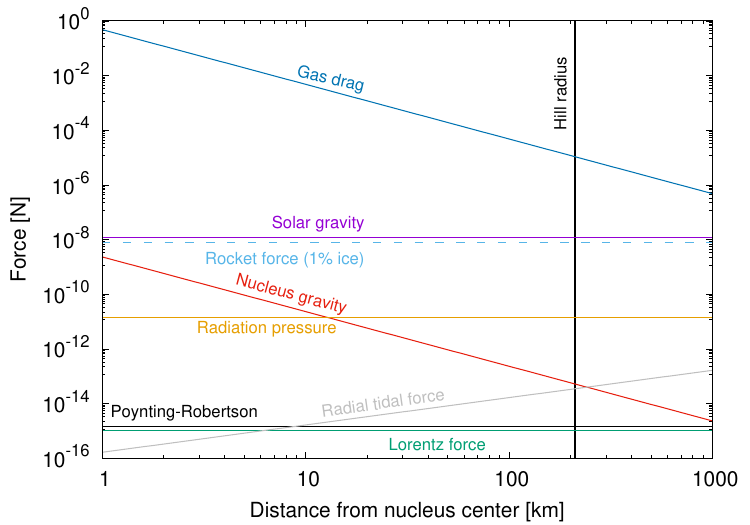}
  \caption{Same as Fig.~\ref{fig:force_vs_size} for fixed particle sizes (left: 10$\mu$m, right: 1mm), 1 au from the Sun, and variable nucleocentric distance. The Hill radius (where tidal force and nucleus gravitational force balance) is indicated by a vertical line.}
  \label{fig:force_vs_distance}
\end{figure*}

\begin{figure*}
  \includegraphics[width=0.49\textwidth]{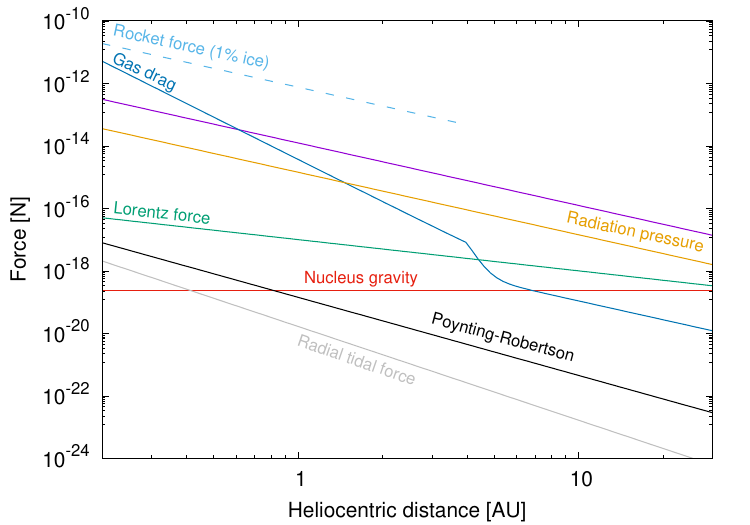}
  \includegraphics[width=0.49\textwidth]{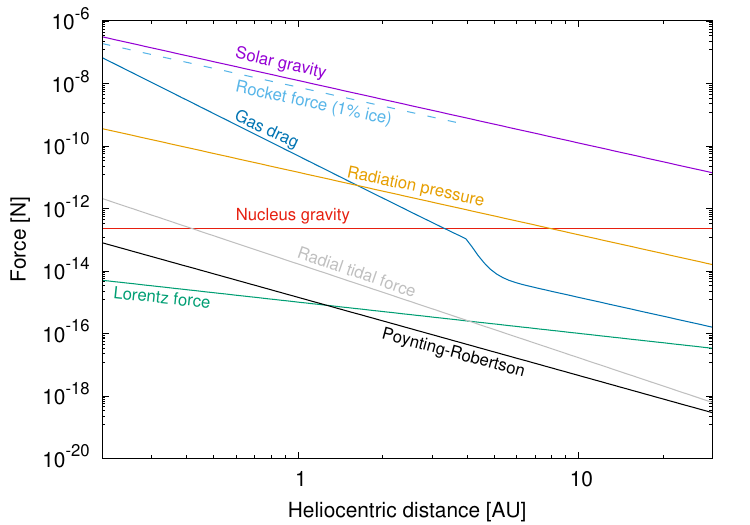}
  \caption{Same as Fig.~\ref{fig:force_vs_size} for fixed particle sizes (left: 10$\mu$m, right: 1mm), nucleocentric distance of 100\,km (hence outside the zone of acceleration by gas drag) and variable heliocentric distance. The gas drag force drops abruptly near 4 au, where we assume a steep drop of the water production rate. For the same reason, we assume that rocket force from intrinsic outgassing of water ice ceases near 4 au. Rocket force is again represented by a dashed line to indicate that both types of particles are too small to retain ice on dynamically significant timescales if mixed with dust.}
  \label{fig:force_vs_rh}
\end{figure*}

The relative importance of the various forces discussed in Sections~\ref{subsec:gravity} -- \ref{subsec:cohesion} depends mainly on the particle size and the distances from the Sun and comet, but also on the dust and gas properties. Figures~\ref{fig:force_vs_size} -- \ref{fig:force_vs_rh} illustrate the key dependencies. 

Nucleus gravity is always several orders of magnitude weaker than solar gravity, but since both the nucleus and the dust are subject to the solar gravitational acceleration, the relevant quantity with which to compare the nucleus gravitational force is the tidal force. The nucleus distance where tidal force and nucleus gravitational force balance is given by the Hill radius.

The Poynting-Robertson effect and the Lorentz force are weakest and therefore typically not considered in calculations of cometary dust dynamics. However, far from the Sun, the Lorentz force on small particles can become comparable to radiation pressure.

If present, outgassing-induced ``rocket'' force tends to be stronger than solar radiation pressure.


Once lifted from the nucleus surface, the initially dominant force on dust particles is gas drag (Eq.~\ref{eq:drag-force}). It decreases roughly quadratically with increasing nucleus distance, as the gas dilutes, and at some distance (that depends on particle size and gas production rate), becomes weaker than solar gravity. We refer to this near-surface regime as the acceleration zone (Section~\ref{sec:acczone}) and to the dust velocity at its upper boundary as the terminal velocity. Once the influences of nucleus gravity and gas drag have diminished, the motion of the dust in the outer coma, tail and trail is largely driven by solar gravity and radiation pressure, and by the terminal velocity the dust acquired from the gas drag. In the outer coma (Section~\ref{sec:outercoma}), dust that was initially ejected towards the Sun from the sunlit surface where water sublimation is strongest, will reverse its direction of motion under the influence of solar radiation pressure, such that eventually almost all dust is driven into the curved tail stretching in the anti-solar direction and along the negative heliocentric velocity vector of the comet, hence outside its orbit (Section~\ref{sec:tailtrail}).

Generally, if particles contain sublimating ice, the forces on them can implicitely be affected (and become time-dependent) by their changing mass, cross section, and temperature. The same applies for fragmentation, spin-changes and changes of temperature.

Often, a maximum liftable grain size is derived from the balance of gravity (Eq.~\ref{eq:grav_nucl}) and gas pressure at the surface (Eq.~\ref{eq:drag-force}). We here formulate it more generally as minimum liftable cross-section-to-mass ratio:

  \begin{equation}
\label{eq:maxlift}
    \left[ \frac{\sigma_{\rm d}}{m_{\rm d}} \right]_{\rm max} = \frac{8 \pi G}{3 C_{\rm D} } \frac{\rho_{\rm N} R_{\rm N}}{m_{\rm g}v_{\rm g}Q_{\rm g}}, 
  \end{equation}
  where $Q_{\rm g} = n_{\rm g} v_{\rm g}$ is the surface gas production rate in molecules per unit time and area. Inserting the values from Table~\ref{tab:dust_parameters} and assuming spherical dust particles, the maximum radius liftable by water vapour alone at 1\,au is about 1\,m. Since gas production rates can be highly variable and hard to measure on a local scale, the actual maximum liftable grain size for a given situation is difficult to predict.

\section{Dynamical regimes}
\label{sec:regimes}

\subsection{Dust emission from the surface}
\label{subsec:surface_emission}

Due to its low optical depth, the surface brightness of dust in the cometary coma is nearly always lower than that of the illuminated surface of the nucleus. With remote sensing methods that measure scattered sunlight or thermal radiation, dust can only be detected against the dark backgrounds of empty space or shadowed surface. It is, therefore, not straightforward to identify the source regions of dust on the surface, even when resolved images of the dust coma obtained by cameras on spacecraft show a considerable fine structure near the limb (cf. Section \ref{subsec:jets}).
Integrated gas production rates indicate that most comets emit only a small fraction of the gas that would be expected from a sublimating surface of pure ice. An exception to this are the so-called hyperactive comets (e.g., 46P/Wirtanen, 103P/Hartley 2, \cite{ahearn11}), in which the global water production is higher than can be explained from pure surface sublimation.

In the following, we describe some of the most common methods used to constrain the distribution of activity across a cometary surface and subsequently outline their findings. A review of local manifestations of cometary activity can be found in \cite{Vincent2019b}.

\subsubsection{Methods to locate activity sources}
\begin{figure*}
  \includegraphics[width=\textwidth]{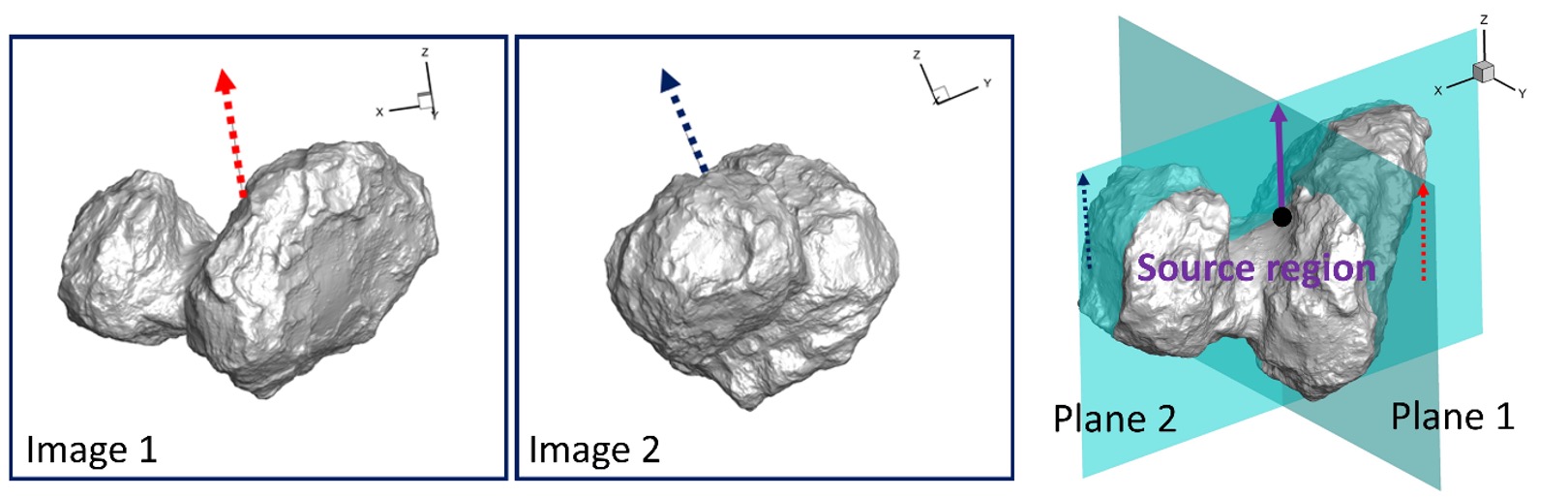}
  \caption{Example of how the source region of a jet can be inferred from images obtained from at least two different perspectives (left and center panels). Each of these two observations renders a plane that contains the line of sight, the jet and its source. The right panel shows how the source region is identified as the point where the intersection line of the two planes crosses the nucleus surface. Image credit: Ian Lai, A\&A, 630, A17, p.3, 2019, reproduced with permission $\copyright$ ESO.}
  \label{fig:lai2019fig3}
\end{figure*}

a) Inversion/triangulation. -- If a bright filament was observed at least twice from different perspectives \citep[ideally 10$^\circ$-30$^\circ$ of sub-observer latitude/longitude, ][]{vincent-oklay2016}, its three-dimensional orientation and source point on the surface can be identified by triangulation: for each image, the projected central line of the filament and the camera position span a plane in three-dimensional space. For two images, the intersection line of the two planes describes the filament axis in three dimensions, and the intersection point of this axis with the nucleus surface (from a shape model) represents the source location (Fig.~\ref{fig:lai2019fig3}). If the filament was observed more than twice, all planes should intersect in the same line within the accuracy limits. This technique is called ``direct inversion'' by \cite{vincent-oklay2016} and relies on the assumptions (1) that the same coma structure can be identified in several images, (2) that at least within a certain distance near the surface, the filament can be described by a straight line, and (3) that its three-dimensional orientation does not change during the time covered by the observations.

Without making the first two assumptions, source locations can still be identified by ``blind inversion'' as long as several images are available. In this approach, the intersection lines of the jet-observer planes with the nucleus surface are calculated for each filament in each image. When combining the surface intersection lines from the different images, lines corresponding to the same filament will intersect in the same point on the surface within the achievable accuracy. Hence the points having multiple intersections are interpreted as the source points of filaments \citep{vincent-oklay2016}. In the presence of many filaments, the results can also be a source density map rather than a map of individual sources. At comet 67P, this inversion technique has been used to trace the origins of both diurnally repeating coma structures by e.g. \cite{vincent-oklay2016,shi-hu2016,lai-ip2019} and of irregular events (``outbursts') by \cite{Vincent2016b}. This technique has also been applied to ground-based coma images \citep[e.g.,][see Section~\ref{subsec:jets}]{farnham-wellnitz2007, vincent-boehnhardt2010, vincent-lara2013}.

If the lower part of a filament is brighter than the background nucleus surface, the source points can also be identified directly \citep[e.g., ][]{agarwal-dellacorte2017,fornasier-hoang2019}.

b) Backtracing of in situ data. --
To constrain the distribution of gas or dust sources across the nucleus surface from in situ measurements aboard a spacecraft, the measured dust or gas density is often used as a proxy for the activity at the sub-spacecraft longitude and latitude at the time of observation \citep[e.g., ][]{hoang-altwegg2017, hoang-garnier2019, dellacorte-rotundi2015, dellacorte-rotundi2016}. The underlying assumption is that the activity changes on timescales long compared to the traveling time of the material from the surface to the detector, and that this motion is radial.
Since this assumption cannot a priori be taken as justified, in particular for the more slowly moving dust, some authors account for the travelling time by including the near-surface acceleration zone when linking the detection coordinates to the ejection point on the surface \citep{longobardo-dellacorte2019, longobardo-dellacorte2020}. These approaches help to understand broad regional and diurnal variations of activity but do not generally provide the spatial resolution to for example link coma material to specific landmarks on the surface.

c) Forward modelling of the motion of dust embedded in the gas flow. --
The source regions of coma dust can be constrained by forward modelling the motion of dust embedded in the gas flow field and iteratively fitting the predicted dust distribution to measurements. Models of the gas dynamics typically either follow a fluid dynamics approach or use the Direct Simulation Monte Carlo (DSMC) method to describe the motion of individual molecules (Marschall et al. in this book).
The comparison of the modelled dust distribution to remote sensing observations requires additional assumptions about the light scattering and/or thermal emission properties of the dust (Kolokolova et al. in this book), and the projection of the three-dimensional dust distribution onto the image plane by line-of-sight integration.
One boundary condition of all coma models is the distribution of gas and dust activity across the cometary surface, which is why they are discussed in the present context. The surface activity distribution needs to be defined such that it enables an optimal reproduction of the data in question. Generally, the obtained solution will not be unique, especially if only the local gas density is used to constrain the models \citep{marschall-liao2020}, but fitting the same model to multiple data sets can reduce the degeneracy.

Forward modelling approaches have been used both to describe temporally and spatially confined phenomema and to understand the global distribution of activity. An example of the former is the study of the influence of topography, local time and illumination conditions on filament structures emanating from the terminator region in \cite{Shi2018}. Global models have for example been used to fit data from the Rosetta Orbiter Spectrometer for Ion and Neutral Analysis (ROSINA), from the Optical, Spectroscopic, and Infrared Remote Imaging System (OSIRIS), and from the Visible and Infrared Thermal Imaging Spectrometer (VIRTIS), and to combinations of such data sets \citep[for example ][]{Marschall2016, Marschall2017, fougere-altwegg2016a, fougere-altwegg2016b, zakharov-crifo2018, combi-shou2020, davidsson-samarasinha2022}.

\cite{kramer-laeuter2017} and \cite{laeuter-kramer2019, laeuter-kramer2020} instead follow an inverse modelling approach, in which the gas emission from each surface element of the shape model is a free parameter, the global distribution of gas in the coma is calculated from the surface production rates, and is probed by the measurements (as a function of space and time) of the ROSINA instrument. The vector of the surface emission rates is connected to the vector of measurements through a matrix and is optimized in order to minimize the deviation between model and measurements. The obtained vector of surface production rates describes the geographical distribution of activity.

d) Torques and non-gravitational forces. --
An additional means of constraining the gas activity distribution across the surface arises from the reaction force that the sublimating gas exerts on the nucleus, similar to a rocket engine. The component of the force crossing the nucleus center of mass leads to an acceleration of the heliocentric orbit, while the component perpendicular to the rotation axis creates a torque that changes the rotation speed and axis orientation. For 67P, a first rough prediction of the change of spin rate during the perihelion passage was made by \cite{keller-mottola2015_rotation}, assuming the local activity to be driven by the illumination conditions and energy balance only. Models aiming to fit simultaneously the rotation state and heliocentric orbit of 67P require a more complex distribution of surface activity \citep{kramer-laeuter2019a, kramer-laeuter2019b, attree-jorda2019, attree-jorda2023}, but a single model to fit both constraints has not yet been identified. A highly asymmetric outgassing has been invoked to explain a rapid decrease in rotation rate of comet 41P/Tuttle-Giacobini-Kres{\'a}k \citep{bodewits-farnham2018}.

e) Surface changes. -- At both comets 9P and 67P, changes on the surface were seen when the same spot was observed multiple times. In particular the Rosetta mission with its two-year coverage of almost the entire surface offered the possibility to study such changes. Surface changes include cliff collapses, receding scarps, formation and extension of fractures and cavities, displacement of boulders, changes in dust mantle thickness and the temporary appearance of bright spots (Pajola et al. and Filacchione et al. in this book). Some of these events, such as cliff collapses and the new exposure of bright surfaces, have been associated to transient dust emission with reasonable certainty \citep{pajola-hoefner2017, agarwal-dellacorte2017}. For fractures and cavities, models have been proposed that connect their formation or deepening to the sudden emission of dust \citep[e.g., ][]{vincent-bodewits2015, skorov-rezac2016}, but observational proof of these models has not yet been found. The connection of moving boulders and scarps and of changes in the thickness of a local dust layer to the emission of dust remains likely but also unproven \citep[e.g., ][]{thomas-ahearn2013, elmaarry-groussin2017, fornasier-feller2019}.

\subsubsection {``Regular'' and ``irregular'' activity}
The appearance of the dust coma and the pattern of filament structures changes with time during a rotation of the comet, but quite accurately repeats with the diurnal cycle \citep{Vincent2016b}. This, together with the observation that most dust is emitted from the illuminated hemisphere \citep{fink-doose2016, gerig-pinzon2020} and that the dust emission follows the subsolar latitude on seasonal timescales \citep{vincent-oklay2016, dellacorte-rotundi2016, lai-ip2019}
indicates that direct solar irradiation is the prime driver of the diurnally repeating (``regular'') activity.

Insolation is, however, not the only factor determining the strength of local dust and gas activity, because models assuming that a constant fraction of solar energy input is consumed by water ice sublimation, and that the dust production is proportional to the outgassing rate, fail to reproduce the in situ measurements of ROSINA, the brightness pattern of coma dust \citep{Marschall2016}, and the non-gravitational forces and torques \citep{attree-jorda2019,kramer-laeuter2019a}. Various patterns of systematic activity enhancements have been proposed, such as enhanced activity from sinkholes \citep{vincent-bodewits2015, prialnik-sierks2017}, cliffs \citep{vincent-oklay2016, Marschall2017}, fractures \citep{hoefner-vincent2017}, and newly illuminated, frost-covered surfaces near receding shadows \citep{prialnik-ahearn2008, desanctis-capaccioni2015, fornasier-mottola2016,Shi2018}. For none of these location types their relative contribution to the global activity has been firmly established. \cite{Vincent2019b} point out that generally, activity from close-to vertical surfaces (having high gravitational slope, i.e. a significant angle between the local surface normal and the negative gravitational acceleration vector), avoids being quenched by a dust mantle, which is an unresolved obstacle to explaining activity from  surfaces with low gravitational slopes.

It is further possible that the activity from ``pristine'' surfaces (cf. Pajola et al. in this book) differs from that originating from terrains that are covered in debris that fell back from the coma. This fall-back material did not reach escape speed when accelerated by the gas and re-impacted at locations where the gas pressure was (at least seasonally) suffiently low \citep{thomas-davidsson2015, pajola-lucchetti2017}. Under conditions of higher seasonal irradiation, this material can again be lifted. This effect has been evoked to explain the strong activity from the neck region Hapi on comet 67P during autumn 2014 that was reported e.g. by \cite{Lin2015, pajola-lee2019, combi-shou2020}. At this time, the comet's approach to the Sun near 3 au led to increased sublimation of water ice, and Hapi, located at high northern latitudes, was in local summer. A significant fraction of the dust cover removed from Hapi during this epoch was later re-supplied, during northern winter in 2015-2016 \citep{cambianica-fulle2020}.

A large-scale trend of enhanced activity from above-average bright and blueish surfaces has been attributed to enrichment of these surfaces with water ice \citep{ciarniello-capaccioni2015, filacchione-capaccioni2016, fornasier-mottola2016, filacchione-capaccioni2020}. On a smaller scale, a direct link between bright, water-ice rich spots on the surface \citep[e.g., ][]{pommerol-thomas2015, barucci-filacchione2016} and local activity has not been established.

In addition to the diurnally repeating activity, many comets show sudden, short-lived events of increased dust emission that are often called ``outbursts''. Such irregular activity has been observed on a wide scale of magnitudes, ranging from small, local-scale dust plumes \citep[e.g., ][]{agarwal-dellacorte2017} to global events easily detectable with Earth-based telescopes \citep[e.g., ][]{lin-lin2009}. 
The processes triggering such events are not well understood, but a wide range of models have been proposed, and the Deep Impact and Rosetta missions have made it possible to study at which locations small outbursts occur.

Comet 67P has shown irregular activity during the whole comet-phase of the Rosetta mission, from April 2014 \citep{tubiana-snodgrass2015} to September 2016 \citep{altwegg-balsiger2017}. The vast majority of these events were not detected from Earth, with the possible exception of one near perihelion \citep{boehnhardt-riffeser2016}. It has been suggested that outbursts occurred mainly near morphological boundaries and cliffs \citep{Vincent2016b, fornasier-hoang2019}, and some were also observed near pits \citep{tenishev-fougere2016} and circular features in the Imhotep region \citep{knollenberg-lin2016, Rinaldi2018, agarwal-dellacorte2017}. One event has been directly linked to the break-off of a cliff face \citep{pajola-hoefner2017}. Temporal concentrations of outburst events have been reported for early morning and local afternoon \citep{Vincent2016b}, but outbursts have also been observed from the deep nightside \citep{knollenberg-lin2016, pajola-hoefner2017, rinaldi-formisano2019}.

The relative contribution of irregular events to the total dust production rate is difficult to estimate, because it depends both on a complete knowledge of their frequency and on the amount of material emitted globally and by outburst events. Estimates indicate that individual events contribute no more than a few percent to the global dust production \citep{tenishev-fougere2016,lin-knollenberg2017}, such that the major part of the dust would be released by the diurnally repeating activity.

\subsubsection{Processes driving the dust lifting}
\label{sec:bottleneck}

Comets or their precursor planetesimals have resided for several billion years in the cold outer solar system: the Jupiter Family Comets in the Transneptunian disc at about 30K, and the Long Period and Dynamically New comets in the, even colder, Oort Cloud at the limits of the Sun's gravitational influence. When an object from one of these reservoirs enters on a trajectory through the region of the planets, the top layers of the surface get heated and begin to lose their volatile ices such as initially CO, and N$_2$ \citep{delsemme1982, laeuter-kramer2019}, then CO$_2$, and, finally, beginning at roughly 5 au from the Sun, H$_2$O. Inside 3 au, water ice sublimation likely becomes the dominant driver of activity, but the more volatile ices like CO$_2$ keep playing an important role in activity \citep[e.g., ][]{ahearn11, combi-shou2020}.

The energy input from solar radiation is partially re-radiated to space and partially absorbed by the porous surface material. The absorbed energy is conducted and radiated to greater depths, where it can cause phase changes in the ices and the release of gas. This gas percolates through porous material, re-condenses on colder surfaces or escapes eventually through the surface. Typically, there is a positive radial temperature gradient: the upper layers are warmer than below. During dusk or in shadowed regions, the temperature gradient may be locally inverted leading to re-condensation of water in the dust mantle. This frost can help to start activity in the morning \citep{desanctis-capaccioni2015, fornasier-mottola2016, Shi2018}.

The global water production of most comets is much lower than expected from freely sublimating ice surfaces. This observation led \cite{kuehrt-keller1994} to conclude that the refractory component in the nucleus must be sufficiently abundant to allow a refractory mantle, depleted from volatiles, to form on the surface and quench activity over wide areas. This model implies that the cohesive Van der Waals forces stabilising the refractory material exceed the gas pressure at the surface by many orders of magnitude, such that the detachment of dust particles from this refractory matrix remains unexplained. \cite{kuehrt-keller1994} suggest that a heterogeneous surface composition might lead to activity of a small fraction of the surface that would have to be considerably enriched in ice.

Laboratory experiments, too, demonstrated that insolation of a porous ice-dust mixture leads to sublimation of the ice, and that part of the gas and dust leave the surface \citep{koelzer-gruen1995}. The non-lifted part of the freed dust builds up a porous dust mantle of about 10 Pa failure stress \citep{gruen-gebhardt1993} that eventually quenches the gas emission. This may be similar on a comet.

Recent estimates of the cohesive strength of cometary material (a few Pa on metre-scales \citep{attree-groussin2018} to kPa inside a dust aggregate \citep{hornung-merouane2016}) and of laboratory analog materials \citep[4-20 kPa, ][]{gundlach-schmidt2018} are indeed larger than the sublimation pressure of water ice in the relevant temperature range (1 Pa at 210K and 10 kPa at 310K).

\cite{gundlach15} point out that overcoming cohesion becomes easier with growing size of the particles constituting the cometary surface, and predict that the size of ejected particles should grow with heliocentric distance. But their given size ranges do not agree well with observed cometary activity, and the model does not explain the activity of very distant comets such as C/2017 K2 that is thought to have onset near the orbit of Neptune \citep{jewitt-kim2021}.

Recent models have tried to overcome the cohesion problem by ascribing a hierarchichal structure to the surface, where the material is organized in ``pebbles'' or aggregates that in turn consist of refractory and optionally H$_2$O and CO$_2$ ice particles (which in themselves may have a substructure and be composed of ``grains''). On the other hand, the ``pebbles'' are clumped into ``chunks'' \citep{gundlach-fulle2020, fulle-blum2020}. The ice content of the pebbles would decrease with time and increase with depth. The porosity of the architecture would prevent the formation of an impenetrable dust mantle but increase sub-surface pressure, while the small number of contact points between pebbles would facilitate to overcome their cohesion. Alternative or additional processes to counteract interparticle cohesion could be related to thermal fracturing or electrostatic charging \citep{jewitt19}.

Reaching a consolidated understanding and consensus on the processes involved in triggering and maintaining the emission of dust from cometary surfaces remains one of the open topics of the field, and is hampered by a lack of constraining data. Laboratory experiments addressing these questions are presented in Poch et al., and the interior structure of comets is discussed in Guilbert-Lepoutre et al., both in this book.

\subsection{Acceleration zone}
\label{sec:acczone}
The dust acceleration region in the general coma extends from the nucleus surface to roughly ten nucleus radii.
At that distance molecule-dust collisions become rare and the dust flow decouples from the gas (Fig.~\ref{fig:dustForcesAndProcesses}).

The brightness in the dust coma is often used as a proxy for the density of dust.
Indeed, when we have an optically thin coma the reflectance, $R$, of dust in a given pixel at a certain wavelength, $\lambda$, and at a certain scattering angle, $\Phi$, is
\begin{equation} \label{eq:reflectance}
  R(\lambda,\Phi) = \int_{a_{\rm min}}^{a_{\rm max}} n_{\rm col}(a) \sigma_{\rm d}(a) q_{\rm eff}(a)  \frac{p(a,\lambda,\Phi)}{4\pi} da,
\end{equation}
where the smallest and largest sizes are given by $a_{\rm min}$, and $a_{\rm max}$, the dust column density along the line of sight is $n_{\rm col}$, and the geometric dust cross-section is $\sigma_{\rm d}$.
The scattering efficiency, $q_{\rm eff}$, and phase function, $p$, depend on the material properties of the dust.
Eq.~\eqref{eq:reflectance} illustrates that the brightness of the dust cannot, in general, be taken as a proxy for the dust density.
There might be a dense part of the coma with particles that have a very low scattering efficiency and thus a low reflectance compared to an area with a smaller number but highly efficiently scattering particles which appear bright.

To understand the structure of the dust coma in the acceleration zone, we will first discuss the radial outflow structure and then go into more detail about how 3D jet-like structures become manifest in the acceleration region.

\subsubsection{The extent of the acceleration region}
\label{sec:extent_acc_reg}
To understand the radial structure of the dust coma let us first consider a simplified coma where the dust is not accelerated but rather flows out radially from the surface with a constant speed.
In this case, the local dust densities decrease with the inverse square of the distance, $r$.
This is due to the fact that mass flux conservation ($n_{\rm d} v_{\rm d} A = {\rm const.}$, where $A$ is the surface area) through closed surfaces around the nucleus is maintained.
If one thinks of these surfaces as spherical shells then their surface areas scale with $A \sim r^2$.
Because the speed, $v_{\rm d}$, is constant in this example, the number density, $n_{\rm d}$, therefore needs to scale with $1/r^2$ for the flux to be constant.

The dust brightness measured by a remote observer is proportional to the column, not the local number, density.
We can use the above behavior and find that the column density will scale as $n_{\rm col} \sim 1/\rho$, where $\rho$ is now the projected distance to the nucleus in the image plane, rather than $r$.

In other words, the column density, and by extension the brightness, $R$, multiplied by $\rho$ is constant for free radial outflow of dust.
This is also the basis for the commonly used quantity $A\!f\!\rho$ \citep{AHearn1984} which -- in the case of free radial outflow -- is independent of where in the coma it is being measured. The quantity $A\!f$ stands for the product of albedo and filling factor (optical depth), and is therefore equivalent to $R$.

\begin{figure*}
  \includegraphics[width=\textwidth]{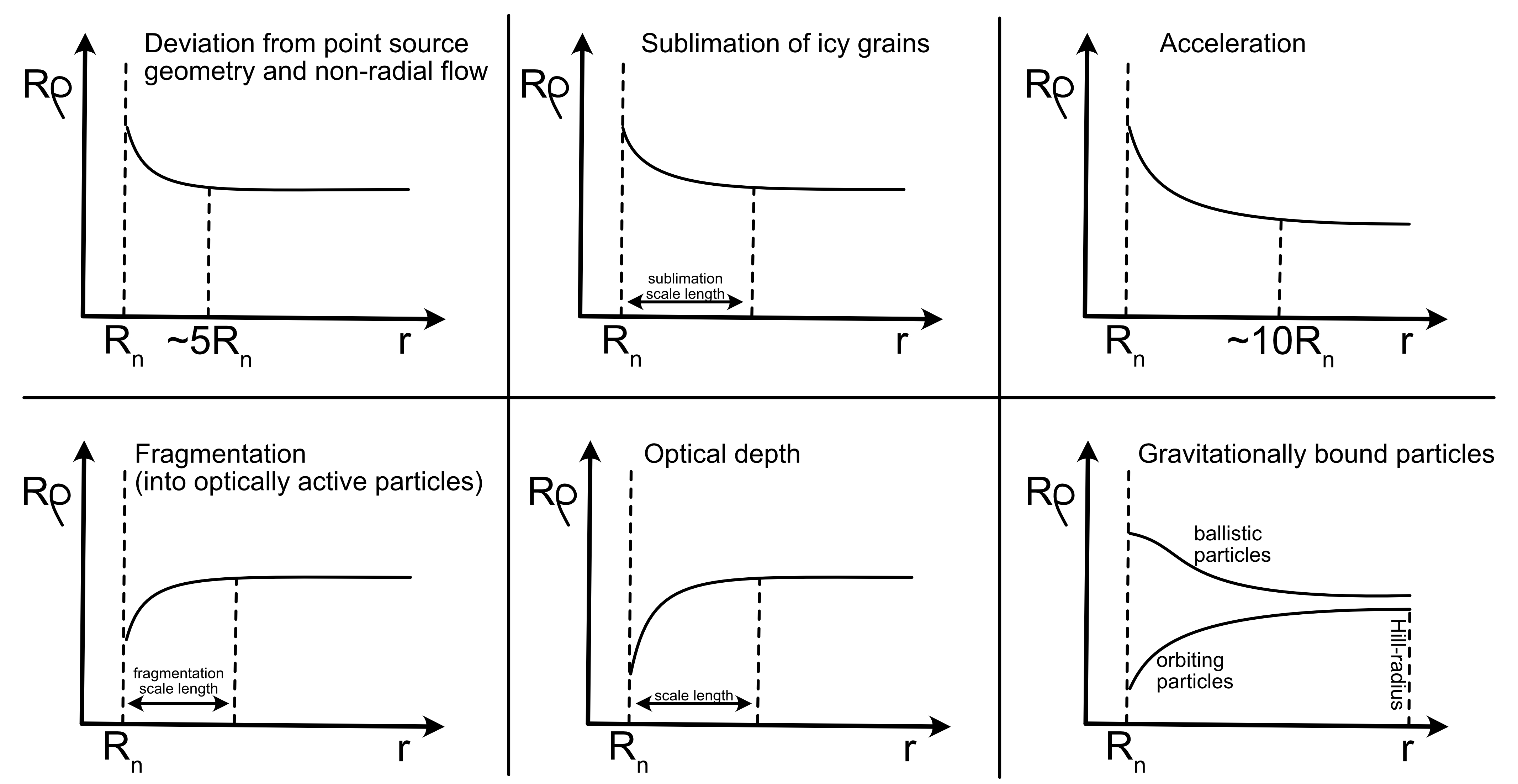}
  \caption{Schematic of how different processes alter the behavior of the product $R\rho$ of dust coma reflectance, $R$, and projected distance to nucleus center, $\rho$, as a function of the distance to the nucleus, $r$. Deviations from a point source and non-radial flow (top left) will reduce $R\rho$ on a length scale of $\sim 5 R_{\rm n}$, where $R_{\rm n}$ is the nucleus radius. The sublimation of particles (top center) will also reduce  $R\rho$ but the length scale of this effect will depend on the properties of the icy particles (their ice content, size), their ejection speed and the heliocentric distance.
        The acceleration of particles (top right) will decrease $R\rho$ and converge to a constant on a length scale of $\sim 10 R_{\rm n}$.
        Both fragmentation (bottom left) and optical depth effects (bottom center) will increase $R\rho$. The scale on which these effects act depends on the details of the processes (i.e., the fragmentation rate).
        Finally, gravitationally bound particles (bottom right) will increase $R\rho$ if they are on bound orbits while decreasing it when on ballistic trajectories. The scale for these gravitational effects is the Hill sphere.
    The figure was adapted from \cite{Gerig2018}.}
  \label{fig:radialOutflow}
\end{figure*}

Here, being interested in the acceleration region, we want to study deviations from a constant value of the product $R\rho$.
Deviation from a constant $R\rho$ can be caused by a multitude of processes.
Sublimation and fragmentation of particles can either in- or decrease $R\rho$ with increasing $\rho$ depending on whether the resulting particles are more or less efficient scatterers (Fig.~\ref{fig:radialOutflow}).
Deviations from a point source nucleus will decrease $R\rho$, optical depth effects will increase $R\rho$, and gravitationally bound particles can increase or decrease $R\rho$ depending on the type of orbit they are on (Fig.~\ref{fig:radialOutflow}).
\cite{ThomasKeller1989} found that the near-nucleus environment of comet 1P/Halley is dominated by optical depth effects. They observed a behavior similar to that shown in the bottom center panel of Fig~\ref{fig:radialOutflow}.
Finally, the acceleration of particles will decrease $R\rho$ (top right panel of Fig.~\ref{fig:radialOutflow}) because as the speed of the particles increases, $n_{\rm d} v_{\rm d}$ decreases more rapidly than with the $1/r^2$ profile described above for free radial outflow.

We can, therefore, use $R\rho$ to determine at which point the dust flow transitions from an accelerated to a free radial outflow, corresponding to the outer edge of the acceleration region.
As this is an asymptotic process there is no hard boundary.
Theoretical calculations \citep{Zakharov2018Icarus} showed the dust particles reach $90\%$ of their terminal speed at around six nucleus radii.
We would therefore stipulate that free radial outflow begins around ten nucleus radii.
This is in very good agreement with observations at and numerical simulations for 67P, where \cite{Gerig2018} found the transition to a constant $R\rho$ at around eleven nucleus radii. \cite{FinsonProbstein1968} derived an upper limit of 20 nucleus radii for the acceleration zone.

\subsubsection{Terminal velocity and transition to the outer coma}
\label{subsec:terminalVel}
As discussed in the previous section, the acceleration region extends to roughly 10 nucleus radii.
This is the interface to the part of the coma where the dust motion is controlled by solar radiation pressure and gravity rather than nucleus gravity and gas drag (Fig.~\ref{fig:dustForcesAndProcesses}).

For numerical simulations, these two force regimes require different algorithmic approaches, such that, in practice, the two parts of the coma are most often treated separately.
Strictly speaking, there is a transition region (Fig.~\ref{fig:dustForcesAndProcesses}) where nucleus gravity still plays a role (the Hill sphere) and large particles that do not reach escape speed will return to the surface.

In any case, it is possible to define a surface -- not necessarily spherical -- where the dust has reached terminal speed and decouples from the gas.
This surface is
the upper boundary of the acceleration region and provides the initial conditions to calculate the dust dynamics thereafter (Sec.~\ref{sec:outercoma}).

In situations where gas drag is the dominant cause of acceleration and where $v_{\rm g} \gg v_{\rm d}$, a simplified dependency of the terminal speed, $v_{\rm ej}$, on particle size, $a$, and global gas number production rate, $Q_{\rm g}$, can be derived analytically. For the given assumptions, Equation~\ref{eq:drag-force} simplifies to
\begin{equation}
  \dot{v}_{\rm d} \approx \frac{C_{\rm D} m_{\rm g}}{2} \frac{\sigma_{\rm d}}{m_{\rm d}} n_{\rm g} v_{\rm g}^2,
\end{equation}
where $m_{\rm d}$ is the dust particle mass. Multiplying with $v_{\rm d}$, assuming purely radial motion with nucleus center distance $r$, describing gas density  as $n_{\rm g}(r) = Q_{\rm g} / ( 4 \pi r^2 v_{\rm g}(r) )$, and integrating from the surface ($r$=$R_{\rm n}$,$v_{\rm d}$=0) to the decoupling distance ($r$=$r_{\rm max}$,$v_{\rm d}$=$v_{\rm ej}$), yields
\begin{equation}
  \label{eq:vej_analytical}
  v_{\rm ej}^2 = \frac{\sigma_{\rm d}}{m_{\rm d}} \frac{Q_{\rm g}}{4\pi} m_{\rm g}  \int_{R_{\rm n}}^{r_{\rm max}}  \frac{C_{\rm D} v_{\rm g}(r)}{r^2} dr.
\end{equation}
The quantities in the integral depend on the radial distribution of the gas speed and on gas and dust temperatures through $C_{\rm D}$. The terminal speed is proportional to the square root of the cross-section-to-mass ratio, $\sigma_{\rm d}/m_{\rm d}$, and equivalently to $\sqrt{\beta}$ or  -- for size-independent density -- to $a^{-1/2}$, and to the square root of the gas production rate, $Q_{\rm g}$.

\begin{figure*}
  \includegraphics[width=\textwidth]{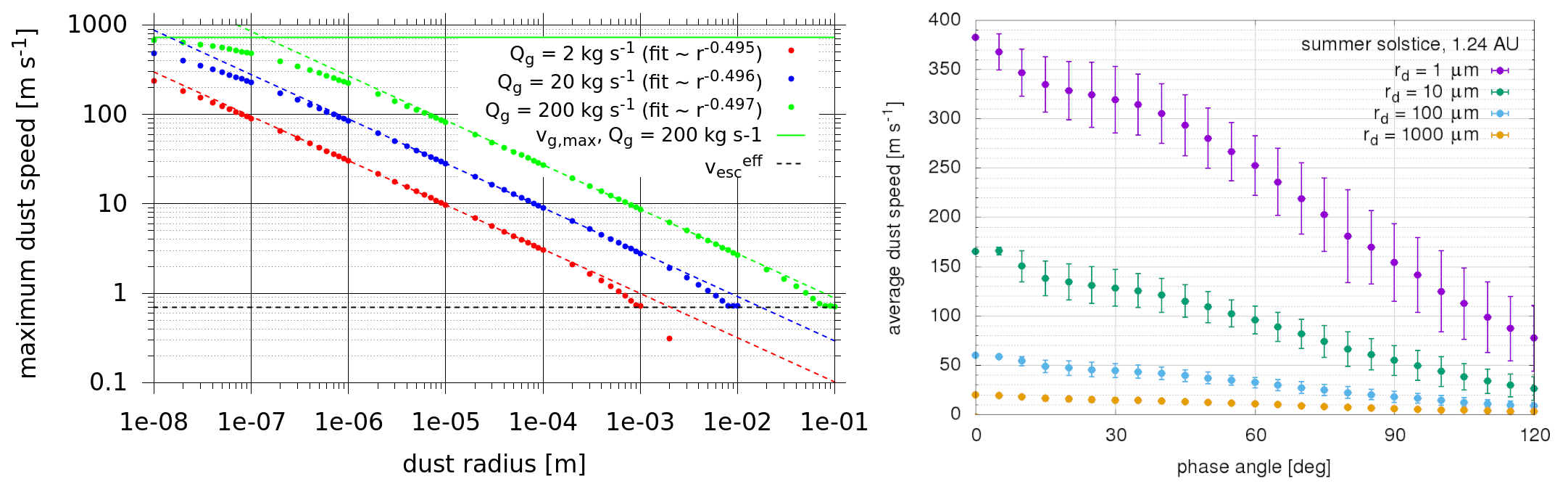}
  \caption{
  The left panel shows the maximum dust speed as a function of dust radius, and gas production rate. The right panel shows the phase angle dependence of the dust speed. Both results shown have been calculated for comet 67P and figures adaped from \cite{Marschall2017PhD}. Speeds $<$2 m s$^{-1}$ for larger-than-centimetre particles have also been reported from 103P/Hartley 2 by \cite{ahearn11}.}
  \label{fig:dustSpeeds}
\end{figure*}

The left panel of Figure~\ref{fig:dustSpeeds} shows the terminal dust speeds as a function of dust size and gas production rate as obtained by numerical simulations.
When the dust is much slower than the gas and the dynamics are dominated by gas drag only then the dust speed scales with $a^{-1/2}$ and $Q_g^{1/2}$, consistent with Equation~\ref{eq:vej_analytical}.
Deviations from this behavior are observed at very small and very large sizes.
A small dust particle accelerates to almost the gas speed and asymptotically approaches it.
The dynamics of large dust particles is significantly influenced by the nucleus gravity and thus their speeds are lower than predicted by Eq.~\ref{eq:vej_analytical}.
Above a certain size, the dust does not reach escape speed and will fall back to the surface. Even larger dust cannot be lifted.

The right panel of Fig.~\ref{fig:dustSpeeds} shows the phase angle dependency of the dust speed.
In the numerical simulations shown there is no night side activity.
Nevertheless, the gas flow and therefore also the dust flow is driven to the night side.
This lateral flow from the day to the night side ensures that even in the absence of night side activity dust particles reach significant speeds at large phase angles.

We would like to re-emphasize that speeds in Fig.~\ref{fig:dustSpeeds} are based on spherical dust particles with the same bulk density as the nucleus of 67P.
If particles are significantly fluffier or rotating, their speeds can increase beyond the values shown in Fig.~\ref{fig:dustSpeeds}.

Dust in locally confined gas sources will attain slower speeds than if accelerated by a global gas field due to lateral gas expansion in the plume, for similar gas production rates per unit area \citep{jewitt_133P}.

\subsubsection{Jets and other 3D structures in the acceleration region}
\label{subsec:jets}
Reducing the acceleration region to the radial expansion would be overly simplified.
Observations from comet 1P/Halley \citep[e.g.][]{Keller1987}, to comet 103P/Hartley 2 \citep[e.g.][]{ahearn11}, and comet 67P \citep[e.g.][]{Lin2015} have shown intricate dust filament structures.

Whether or not these filaments, also referred to as "jets" \citep[see longer discussion in ][]{Vincent2019b}, have clear source regions on the surface
is a critical question.
Three plausible mechanisms can result in the observed filaments:
\begin{enumerate}
\item jets with a clearly defined and confined gas and/or dust source on the surface;
  \item topographically sculpted filaments that are products of the local topography shaping the dust emission through self-shadowing and/or the underlying convergence of the gas flow;
  \item optical illusions, originating from a large area on the surface and appearing as narrow structures only from specific viewing geometries.

\end{enumerate}

The first mechanism encompasses outbursts or exposed icy surfaces with enhanced activity compared to the background.
Many outbursts have been spatially and temporally resolved at e.g. 67P \citep[e.g.][]{Vincent2016b,Rinaldi2018}.
They are characterized by a sudden increase of the dust emission, peaking after a few minutes, followed by a smooth decrease of the emission to the pre-outburst level.
These events are among the few situations that can with high certainty be characterized as "jets" or "plumes" in the strictly physical sense \citep[see also][]{Vincent2019b}.

\begin{figure}
  \includegraphics[width=0.5\textwidth]{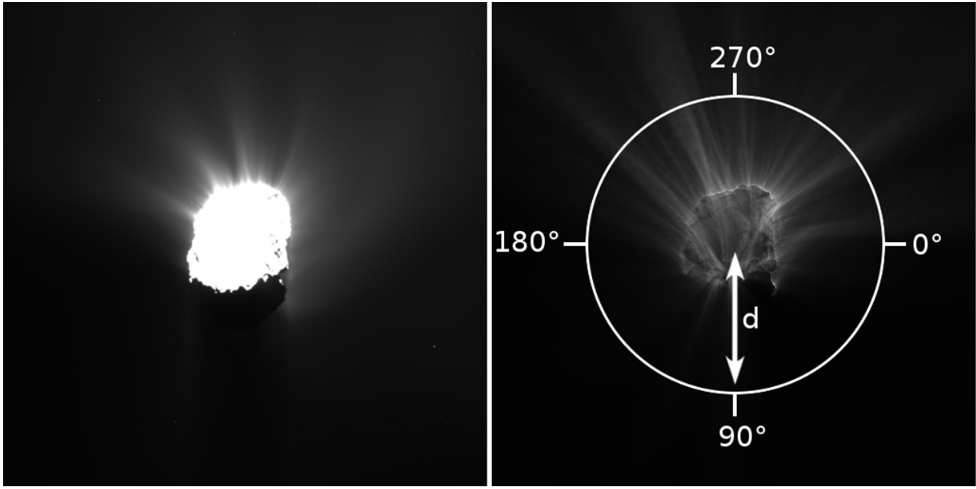}
  \caption{Left panel: Stretched and cropped OSIRIS-WAC image of the coma of comet 67P taken on 2015-05-05 06:28:54 UTC. Right panel: Modeled dust brightness from \cite{Marschall2019} but the nucleus is not overexposed as in the actual OSIRIS image.}
  \label{fig:filaments}
\end{figure}

The second mechanism is related to the irregular topography of the surface.
\cite{Crifo2002} and \cite{Crifo2004} have pointed out that dust structures in the coma do not require sources on the surface.
Non-spherical nucleus shapes are sufficient to produce such features dynamically.
The uneven surface topography focuses the gas and hence the dust flows, resulting in higher density regions within the coma.
More complex nucleus shapes can produce more intricate structures in a total absence of localized sources.
For 67P, this was illustrated in \cite{Marschall2016,Marschall2017,Marschall2019} and Fig.~\ref{fig:filaments}.
This work would indicate that the dust filaments observed in the coma of comet 67P do not have a source area in the traditional sense.

The third mechanism listed above is essentially a mirage.
A prime example is the big fan-like structure originating from the northern neck of 67P.
\cite{Shi2018} demonstrated that this feature likely originates from a sublimating frost front on the morning terminator deep in the neck of 67P.
The out-flowing dust particles produce a kind of fan that, seen from perpendicular to the plane of the fan, is indistinguishable from the surrounding coma.
But with the line of sight inside this plane, the dust column densities in the projected fan are high in contrast to the surrounding coma. The resulting fine filament structure, however, is partly an optical illusion rather than a ``jet'' in the physical sense.

It appears that optical illusions and topographic sculpting are the rule rather than the exception to explain filament structures in the near-nucleus environment \citep{Shi2018, Marschall2019, Vincent2019b}. The current state-of-the art modeling suggests that no confined sources of these features are required. Rather the much simpler assumption of a mostly homogeneous surface, the topography, and viewing geometry are sufficient to explain the filamentary inner coma environment.

\subsection{Outer coma}
\label{sec:outercoma}

The outer dust coma begins outside the acceleration zone and ends at the tail regime.  These distances will vary by grain parameters, but for most active comets with grain radii $\lesssim 1$~mm the outer coma spans from nuclear distances of a few nucleus radii to of order 10$^4$ km.  The dominant forces acting on the outer coma are nucleus gravity, solar gravity, and radiation pressure.  In the absence of grain fragmentation or outgassing, the fine-grained dust coma is typically in a radial outflow.  Thus, the grain number density varies with $r^{-2}$, which produces the canonical $\rho^{-1}$ coma in telescopic observations, where $\rho$ is the projected distance to the nucleus (Section~\ref{sec:extent_acc_reg}).  Within the Hill sphere, however, large particles may be bound to and orbiting the nucleus.

\subsubsection{Coma size-$\beta$-speed relationship}
\label{sec:turn-back}
The size of the dust coma depends on the physical properties of the grains (size, mass, and optical properties), and the ejection speeds imparted on the dust by gas pressure in the inner coma, but also from fragmentation processes or intrinsic outgassing, if relevant.  In the simple case, solar radiation pressure is the primary non-gravitational force accelerating the dust grains.  The acceleration is continuous, so there is no trivial delineation between the coma and tail regimes.  One commonly adopted parameter is the apparent turn-back distance, $X$, which is the distance at which a grain ejected directly at the Sun will reach a speed of 0 in the rest frame of the nucleus. It is derived by integrating the equation of motion of a grain accounting for radiation pressure, ejection speed, and projection onto the sky:
\begin{equation}
  X = \frac{v_{\rm ej}^2 \sin{\theta}}{2 a_{\rm r}} =
  \frac{r_{\rm h}^2 v_{\rm ej}^2 \sin{\theta}}{2 G M_\odot{} \beta} =
  \frac{8 \pi a\rho c r_{\rm h}^2 v_{\rm ej}^2 \sin{\theta}}{3 \mean{Q}_{\rm pr} L_\odot},
  \label{eq:turn-back}
\end{equation}
where $v_{\rm ej}$ is the terminal speed of the dust grain after leaving the gas acceleration zone, $a_{\rm r}$ is the acceleration from radiation pressure ($F_{\rm r}/m_{\rm d}$; Eq.~\ref{eq:radpr}), $\theta$ is the phase (Sun-comet-observer) angle, and $\beta$ is defined by Eq.~\ref{eq:beta}. Equation~\ref{eq:turn-back} is relevant for heliocentric distances $\gtrsim4R_\odot$ \citep{lamy74-dust}.  For isotropic expansion, the turn-back distance traces a paraboloid on the sky \citep{michel76-dust-coma}.

Based on Eq.~\ref{eq:turn-back}, we can estimate an order of magnitude coma upper-limit size. For $v_{\rm ej}=200$~m~s$^{-1}$, $a=1$~\micron{}, $\rho=500$~kg~m$^{-3}$, and $\theta=90$\degr{} (i.e., no foreshortening due to phase angle), $X=10,000$~km at 2~au from the Sun.

However, $v_{\rm ej}$ is a function of gas production rate and dust properties.  In Section~\ref{subsec:terminalVel}, we showed that when the dynamics are dominated by gas drag, dust terminal speeds tend to scale with $a^{-1/2}$ for spherical grains of constant density.  Under these assumptions, $X$ is approximately independent of $a$, but scales with the constant that relates $v_{\rm ej}^2$ to $a^{-1}$, or alternatively with $Q_{\rm g}$.

The dependence of coma size on heliocentric distance is mainly given by the factor $r_{\rm h}^2 Q_{\rm g}(r_{\rm h})$, where there is no strong consensus regarding the shape of $Q_{\rm g}(r_{\rm h})$. It is often approximated by a power law $r_{\rm h}^k$ at least in confined intervals of $r_{\rm h}$. Calculating $Q_{\rm g}$ from the balance of sublimation, thermal radiation and solar irradiation on a perpendicularly illuminated water ice surface gives $Q_g \propto r_{\rm h}^{-2}$ inside 1 au (where radiation cooling is negligible compared to sublimation cooling). Beyond, the exponent, $k$, of a locally fitted power law transitions smoothly to $k=-$4 at 4 au, and becomes even steeper beyond. However, beyond 2 au, the sublimation of more volatile species makes a relevant contribution to the total gas production rate, flattening its heliocentric profile. Observations suggest a much steeper than $r_{\rm h}^{-2}$ profile for water inside 3.5 au: for 67P, \citet{hansen-altwegg2016} find $k=-$5.3 before and $k=-$7.1 after perihelion, while \citet{marshall-hartogh2017} find $k=-$3.8 before and $k=-$4.3 after perihelion. 
Generally, the exponent $k$ seems to be smaller than --2 in most situations, and hence, coma size should rather grow with decreasing heliocentric distance.

The dependency of $X$ on $r_{\rm h}$ is further -- through $v_{\rm ej}$ -- affected by the dependency of $v_{\rm g}$ on $r_{\rm h}$. Various approximations exist for $v_{\rm g} (r_{\rm h})$: $v_{\rm g} \sim r_{\rm h}^{-0.5}$ \citep{tseng07}, $v_{\rm g} \sim r_{\rm h}^{-0.4}$ within $r_{\rm h}$ = 7 au and $v_{\rm g} \ne v_{\rm g}(r_{\rm h})$ beyond that distance \citep{biver-bockelee2002}, while hydrodynamic model calculations by \cite{muellerPhD} suggest weak dependence of $v_{\rm g}$ on $r_{\rm h}$.

Rather than deriving coma-size, Eq.~\ref{eq:turn-back} or related considerations are often applied to the measured coma size in order to estimate coma grain properties from assumptions on velocity and/or acceleration. 
A few recent comets can serve as examples of the variety of conclusions that can be drawn for this analysis.

\citet{jewitt19} measured a growth in the $\rho^{-1}$ coma size of comet C/2017~K2 (PanSTARRS), and use it to estimate a timescale of activity.  Combining the magnitude of radiation pressure acceleration with this timescale, they estimate from the absence of a detectable tail that the optically dominant dust grains must have $\beta>0.003$. The paucity of small grains in the coma suggested the influence of particle cohesion, which prevents their release and favours the production of large particles \citep{gundlach15}.

\citet{hsieh04} estimated ejection speeds $\ll45$~m~s$^{-1}$, based on the lack of a resolved coma in images of 133P/Elst-Pizarro and an assumption of small, $\beta=1$ particles.  \citet{mueller13} measured the growth of a narrow dust feature in images of comet 103P/Hartley~2.  They found that the source was most likely active for $\approx22$~hr, longer than the long-axis precession of the nucleus, which had consequences on their derived source location for the feature. Finally, \citet{kelley13} studied point-sources in the coma of comet 103P/Hartley~2, and with Eq.~\ref{eq:turn-back} concluded that the dynamics of these $\gtrsim1$-cm sized particles were not governed by radiation pressure.

\subsubsection{Large particles in the coma}

Large particles or chunks of nucleus, i.e., centimeter-sized and larger, may be ejected from the nucleus or inner coma with very low speeds, and potentially placed into sub-orbital trajectories.  Under the influence of an additional force, the particles can be placed into bound orbits around the nucleus.  The force may arise from gas outflow anisotropies in the inner coma, or from outgassing of the large particles themselves.

Evidence for centimeter-sized and larger particles may be found in cometary dust trails, meteor showers associated with comets, and in observations of comets at sub-millimeter to centimeter wavelengths, including radar.  See Ye et al. in this book for a review of cometary meteor showers, and \citet{harmon04} for a review of radar observations of cometary comae.  Dust trails are addressed in Section~\ref{subsec:zodiacal}.

Observations of individual particles in the coma, including those in bound orbits, are a more recent phenomenon.  \citet{ahearn11} presented images from the Deep Impact spacecraft of the inner coma of comet 103P/Hartley 2 containing thousands of point sources within a few kilometers from the nucleus.  Such particles were not reported at 9P/Tempel~1.  Depending on the light scattering properties of the particles, they may be as large as meter-sized.  The presence of such large nuclear chunks is a potential solution to the comet's hyperactivity.
The large chunks provide additional sublimating surface area, which enhances the water production rate of the comet \citep{kelley13,kelley15,belton17}.

Point sources were also seen by the Rosetta spacecraft upon its approach to comet 67P/Churyumov-Gerasimenko.  \citet{rotundi15} estimated their sizes, assuming nucleus-like properties, and found the largest to be meter-sized.  The particles are likely in bound orbits, and remnants from the comet's last perihelion passage.  Particles seemed to fill the Hill sphere at the time of the observations (radius 318~km at 3.6~au from the Sun).

Outgassing of the large particles seems to be an important dynamical process, at least for those that are freshly ejected from the nucleus.  \citet{kelley13} showed that comet 103P/Hartley 2 had an asymmetry in its near-nucleus ($\lesssim10~$km) population of large particles, and concluded that acceleration by outgassing best accounted for their distribution and their speeds as measured by \citet{hermalyn13}.  These particles were also likely responsible for Hartley~2's OH-tail observed by \citet{knight13} and the tailward enhanced rotational temperature seen by \citet{bonev13}.

\citet{agarwal16} observed the acceleration of decimeter-sized particles in the vicinity of the nucleus of comet 67P/Churyumov-Gerasimenko.  The acceleration was not strictly in the anti-sunward direction.  They concluded that acceleration from outgassing and the ambient coma were the processes most likely to be responsible for the observed motions.

On larger spatial scales, particle outgassing may be less important.  \citet{reach09} argued that on the basis of the width of the trail of comet 73P/Schwassmann-Wachmann 3, trail particles $<10$~cm in radius are either ice-free after ejection from the nucleus, or are quickly devolatilized.

A fraction of the particles or chunks in the centimetre-to-decimetre size class fall back to the surface where they form smooth layers of fallback material that was observed to cover wide regions of comet 67P. \cite{Marschall2020} estimate that between 11\% and 22\% of the debris mass initially lifted off the surface falls back. The fallback material likely still contains substantial amounts of water ice \citep{davidsson-birch2021}.

\subsubsection{Connecting spacecraft and remote observations of comae}
\label{subsec:sc_and_remote_comae}

If the regular dust structures in the acceleration region do not have local sources in the canonical sense (Section~\ref{subsec:jets}) the question arises whether structures observed in the outer coma can be traced back to the surface or not. Generally, from telescope images of outer coma jet features alone, the physical processes causing these features cannot be inferred. But telescopic images have been used to infer the properties of cometary nuclei, including rotational state and number and distribution of active areas \citep{Vincent2019b}. These results often rely on the presence of distinct jet-like features in the data.

It has been demonstrated that one can reliably trace large scale dust coma structures back to a virtual surface that would be the outer edge of the acceleration zone. It seems possible to expand this inversion down to the nucleus surface at the cost of increased spatial uncertainty on the source location, if assuming that the emission is on average perpendicular to the surface and that the emission vector measured at the edge of the acceleration zone is essentially a weighted average of all contributions. From a purely geometrical point of view, the angle between the measured emission vector and the nucleus north pole defines the effective co-latitude of the source on the surface (accounting for large scale topography). This technique was successfully applied by Vincent et al. (2010, 2013) and Farnham et al. (2007, 2013) to infer the location of specific sources on the surfaces of comets 9P, 103P, and 67P from ground-based observations alone. The inverted source locations were confirmed by in-situ measurements from Deep Impact and Rosetta.

Hence, the connection between dust coma morphology in ground-based observations may be further investigated when in situ spacecraft observations are available for comparisons.

  A rich set of dust features was observed in the coma of comet Halley during its 1986 perihelion passage, including shells, arcs, and nearly linear features, which were also observed in previous apparitions \citep{larson-sekanina87}.  \citet{sekanina-larson86} interpreted the nearly linear features in ground-based images as repeated discrete ejection events, due to their alignment with dust synchrones.  Images of the comet from the Giotto spacecraft \citep{Keller1987} show regions of strong activity from small, approximately kilometer-sized regions \citep[nucleus dimensions are $7\times7\times15$~km;][]{merenyl90-halley}.  The model of \citet{belton91}, with 5 localized active areas on a nucleus with an excited spin state, successfully combined coma features seen in ground-based data with the inner-coma jets observed by the Giotto and Vega spacecraft.

  The next inner coma and nucleus to be imaged by spacecraft was comet 19P/Borrelly.  The comet has a prominent asymmetry due to a jet-like feature in ground-based data \citep{farnham02}.  This feature is not aligned with the sunward or expected dust tail directions, indicating it is produced by directed emission rather than radiation pressure effects.  Deep Space 1 images show that this jet-like feature is due to topography \citep{soderblom04}.  This long bi-lobed nucleus has a rotational axis perpendicular to the long axis of the nucleus, and the telescopically observed jet is due to dust released from its long flat polar region.  A similar effect was seen from the flat southern polar region of 9P/Tempel 1 \citet{farnham-wellnitz2007}.

  The inner coma of 81P/Wild 2, imaged by the Stardust spacecraft, presented many jet-like features and filaments \citep{sekanina04}. Ground-based images have shown two prominent dust coma asymmetries: (1) a broad fan directed to the north of the orbital plane, mainly active pre-perihelion; and, (2) a narrow jet-like feature, directed more sunward and to the south of the orbital plane, mainly active post-perihelion \citep{sekanina03}.  The Stardust flyby was 98 days after perihelion, and source (1) was not active at the time.  Source (2) should have been active, but \citet{farnham05} could not single out any of the Stardust-observed filaments as candidates for its source.  They concluded that source (2) might have been temporarily inactive due to the diurnal rotation of the nucleus, or is the result of a combination of several filaments.  This highlights one of the potential issues with remotely-observed dust features.  As compared to gas, dust coma expansion speeds tend to be low, $\lesssim100$~m~s$^{-1}$, and a broad range of ejection velocities may be imparted on the grains (e.g., Fig.~\ref{fig:dustSpeeds}).  Thus material propagating outward from an active source on a rotating nucleus might trace out an arc or partial spiral for a unimodal speed distribution, but would be less pronounced, blurred, or lost altogether for a broad speed distribution \citep[e.g.,][]{samarasinha00}.

  In addition to the broad southern feature discussed above, enhanced ground-based images of comet 9P/Tempel 1 show other dust coma features \citep[e.g.,][]{lara06}.  \citet{vasundhara09} used a spherical nucleus model and ground-based and Deep Impact data to derive four effective dust sources at the nucleus.  \citet{vincent-boehnhardt2010} used a nucleus shape model based on Deep Impact flyby data to determine source locations for the features, finding six active regions in total with some compatibility with the \citet{vasundhara09} results.

  Deep Impact spacecraft data of comet 103P/Hartley 2 show a strong active area located on the small end of this bi-lobed nucleus \citep{ahearn11}.  Dust coma features in ground-based data were observed, e.g., by \citet{mueller13} and \citet{lin13}.  However, the nucleus is in an excited rotation state, which complicates connecting coma features seen remotely to the inner-coma features and the nucleus seen by Deep Impact.  Regardless, \citet{mueller13} did compare images taken contemporaneously to the Deep Impact spacecraft's closest approach to the comet, and proposed that two secondary features in the coma originated from active areas found along the long axis of the nucleus and near the solar terminator at the time of the Deep Impact flyby.

  In contrast with all previous cometary spacecraft missions, the Rosetta mission enabled a much broader comparison to ground-based data, owing to its long ($\sim$2~yr) residence near the nucleus.  Furthermore, the comet's apparent orbit-to-orbit stability in terms of activity \citep{snodgrass17} enables comparisons beyond the 2015/2016 apparition.  \citet{vincent-lara2013} studied the morphology of ground-based images of the dust coma in the 2003 and 2009 apparitions, and derived a pole orientation and the planetocentric locations of three active areas.  Furthermore, \citet{knight-snodgrass2017} found good agreement with their dust coma observations and the predictions of \citet{vincent-lara2013}.  They indicated one active area was the Hapi region (the ``neck'' of the bi-lobed nucleus), but concluded the other two were less obvious, with one possibly connected to the southern region, and the other to Imhotep (a flat smooth region on the largest lobe).

\subsubsection{Dust size distributions}
The size or mass distribution of dust describes the relative abundance of particles of different sizes or masses, and is usually approximated by power-laws for defined size (mass) intervals, which includes ``broken'' power-laws that have different exponents for different size (mass) ranges. The employed power-laws describe either the differential or the cumulative distributions. The conversion between their respective exponents, and between size and mass distributions is given by the rules of differential calculus. For such conversions, usually the assumptions of size-independent bulk density and spherical particle shape are made \citep[e.g., ][]{agarwal-mueller2007}.

The exponent of the power-law determines whether the largest or smallest particles in the concerned interval dominate the mass and scattering cross-section of the particle ensemble. This exponent cannot only change with size, but also with a comet's position relative to perihelion \citep{fulle-colangeli2010, fulle-marzari2016, merouane-stenzel2017}. Different measurements, being sensitive to different types of particles, can yield different exponents as well \citep[e.g., ][]{blum-gundlach2017, rinaldi-dellacorte2017}. 

The size distribution of dust observed in the tail or outer coma will not generally correspond to size distribution of dust lifted from the surface (due to back-falling or orbiting particles), and even less to that of material resting on the surface (e.g., due to particles not liftable, Eq.~\ref{eq:maxlift}) or inside the deep interior. The size distribution of escaping dust may further be affected by fragmentation and sublimation of a potential volatile component.

The dust size distribution in a comet may carry information about its building blocks and formation process, provided that any post-formation changes to that distribution are understood and accounted for.

The unknown fraction of fall-back material also complicates attempts to infer the refractory-to-ice ratio in a cometary interior, even when the masses of the escaping dust and gas are known, as -- at least integrated over a whole perihelion passage -- is the case for comet 67P \citep{choukroun-altwegg2020}.

It is possible that the dust size distribution, and especially the particle size containing most of the light scattering cross-section, varies between comets. For some prominent, bright, long-period comets, the particles dominating the interaction with light could be micron-sized. Assumed indicators of this are the presence of a strong silicate emission feature at wavelengths near 10~\micron\ and a high maximum degree of linear polarisation of the scattered light \citep{kolokolova-hanner-cometsII-2004}, although similar characteristics are also expected from aggregates of (sub-)micrometer grains. \cite{fulle-cometsII-2004} reports particularly a dominance of micron-sized particles in the scattering cross-section of long period comets Hyakutake and Hale-Bopp, but also in the active Centaur Chiron. Dust instruments flying by comet 1P/Halley detected micron-sized dust with a size distribution that makes it the optically dominant component of Halley's dust \citep{mcdonnell-green1989}. Also at comets 81P/Wild 2 \citep{green-mcdonnell2004} and 9P/Tempel 1 \citep{economou-green2013} micron-sized dust was detected by in situ instruments during flybys. Micron-sized particles were also among those returned from comet Wild 2 to Earth by the Stardust spacecraft \citep{brownlee2014}.

In other comets, and measured with other methods, the main scattering cross-section seems to rather be in (sub-)millimeter-sized grains. One indicator is the absence of a prominent radiation-pressure swept tail in distant comets C/2014 B1 \citep{jewitt_B1_2019}, C/2017 K2 \citep{jewitt19} and the interstellar comet 2I/Borisov \citep{kim-jewitt2020}. \cite{reach07} investigated thermal infrared emission from the debris trails of some 30, mainly Jupiter-family comets and found that in most of them, the amount of millimeter-sized dust required to explain the brightness of the trails was also sufficient to explain the scattering cross-section in the coma, such that not much scattering should have been contributed by potential additional micron-sized particles in the coma.

\begin{figure}
  \includegraphics[width=0.5\textwidth]{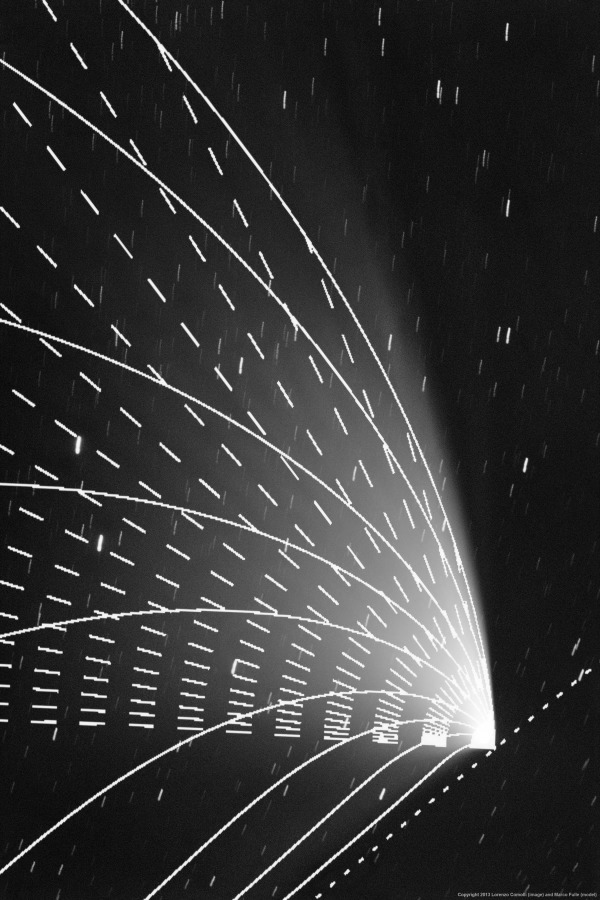}
  \caption{Synchrone-syndyne network for Comet C/2011 L4 (PANSTARRS) on 2013 March 21. The short near-vertical lines are trailed background stars. Image Credit: L. Comolli - Model Overlay: M. Fulle.}
  \label{fig:synsyn}
\end{figure}

\subsection{Tail, trail and dispersion into the zodiacal cloud}
\label{sec:tailtrail}
\subsubsection{Key parameters}

The tail regime begins outside the Hill sphere (Fig.~\ref{fig:dustForcesAndProcesses}).
The trajectory of a dust particle in the tail region and beyond is determined by the particle size (through the radiation pressure parameter $\beta$ according to Equation~\ref{eq:beta}) and the ejection terminal velocity $v_{\rm ej}$ \citep{FinsonProbstein1968}.

Most dynamical models use a simplified treatment of comet dust assuming $Q_{\rm pr}$ = 1. A more detailed treatment dependent on dust mineralogies and structures (compact or fluffy) is discussed in Kolokolova et al. in this volume.

The generally applicable form for particle ejection by gas drag is given by \citep[][see also Section~\ref{subsec:terminalVel}]{whipple1950}:

\begin{equation}
  v_{\rm ej} \propto \beta^{1/2} .
\end{equation}

After a cometary dust particle has been ejected from a nucleus and left the dust-gas coupling region (Fig.~\ref{fig:dustForcesAndProcesses}), its motion is mainly controlled by solar gravity and radiation pressure.
The size distribution of cometary dust has been inferred from both remote sensing images and in situ data.
In general, it is assumed that the distribution of particle radii can be approximated by a power law so that the number of particles with radii ranging from $a$ to $a+da$ is $n(a) da = \Gamma a^{-\alpha}da$.  Table (\ref{size}) provides a summary of $\alpha$ values reported in recent literature.
The mean $\alpha$ values in Table (\ref{size}) is $\alpha$ = 3.7$\pm$0.2 and the median is $\alpha \sim$ 3.4.
Many comets in the past were characterized by $\alpha$ = 3.5, typical of particles in collisional equilibrium \citep{dohnanyi1969}.

\subsubsection{General landscape of a dust tail}

Assuming that $v_{\rm ej}$ = 0, the loci of particles of a given $\beta$ and different ejection times are defined as a syndyne curve \citep{FinsonProbstein1968}, and the loci of particles with different $\beta$ but the same ejection time are defined as a synchrone curve. A specific example of a synchrone-syndyne network is shown in Figure~\ref{fig:synsyn}. An online tool is available for generating synchrone-syndyne diagrams \citep{Vincent2014}\footnote{\url{https://comet-toolbox.com/FP.html}}.
This two-dimensional model is used for simple analysis of comet tail morphology and has been employed to determine the $\beta$ range and ejection times of dust, including by sporadic emission events.
In reality, $v_{\rm ej}$ is not zero. This leads to an expansion of the syndynamic tube whose width is given by the dust ejection velocity.
\citet{fulle-cometsII-2004} pointed out that syndyne analyses tend to yield misleading $\beta$ values, and thus, a three-dimensional dynamical model is needed to consider non-zero ejection velocities.

Cometary dust tails, especially when dust sizes are small, can strongly interact with the solar wind, including the disruption or disconnection of the tail \citep[e.g.,][and Section~\ref{sec:em_forces}]{jones-knight2018}.

\subsubsection{Tail dynamical modeling}
Tail dynamical modeling is a useful technique to explore the properties and ejection processes of cometary dust, which are frequently underconstrained by observations.
Several authors have developed various three-dimensional dynamical models to match the observed images
\citep[e.g.,][]{fulle1989, lisse-ahearn1998, reach-sykes2000, ishiguro-sarugaku2007, moreno2009}. The description of the underlying theory in \cite{fulle-cometsII-2004} is still valid. We here present a simplified model as a guide.
For simplicity, we do not consider the heliocentric distance dependence on the ejection velocity and the dust production rate.
The ejection terminal velocity of dust particles is given by
\begin{equation}
  v_{\rm ej} = V_0 \beta^{1/2},
\end{equation}
where $V_0$ is the mean ejection speed of particles with $\beta= 1$.
Particles are assumed to be released from the sunlit hemisphere of the nucleus.
Assuming that the particle size follows a simple power-law with an index $\alpha$, the dust production rate for a given size and time can be expressed as
\begin{equation}
  N(a_{\micron},t)~da = N_0 a_{\micron}^{-\alpha} da,
\end{equation}
in the size range of $a_{\rm min} \leq a_{\micron} \leq a_{\rm max}$.
The mean dust production rate of 1~$\micron$ particles, $N_0$, can be determined by comparison with the calibrated images.
Model images are generated using Monte Carlo simulations by either solving Kepler's equations or using N-body integrators that include solar gravity and radiation pressure, substituting the gravitational constant $G$ by $G(1 - \beta)$ \citep{chambers1999, ye-hui2014}.
The cross-sectional area of dust particles in the CCD pixel coordinate system by integrating over time and particle size is given by
\begin{equation}
  C(x,y) = \int_{t_0}^{t_1} \int_{a_{\rm min}}^{a_{\rm max}} N_{\rm cal}(a_{\micron},t,x,y) \pi a_{\micron}^2 da_{\micron} dt ,
\end{equation}
where $N_{\rm cal}(a_{\micron}, t, x, y)$ is the number of particles projected within a pixel of the CCD image.
The modeled image is convolved with a Gaussian function whose FWHM equals the FWHM of the seeing disc or pointspread function of the telescope image as applicable, and the resulting image is compared with the observed image to find a range of possible parameters and infer the underlying ejection process.

\subsubsection{Comparison between different types of comets}

There has been extensive modeling of dust emissions from different types of comets, such as Jupiter-family comets, long-period comets, main-belt comets, and interstellar objects. As in the outer coma regime (Section \ref{sec:outercoma}), the main scattering cross-section in the tail regime seems to be in sub-millimeter and larger particles, which were the result of continuous emission that occurred several months to years prior to the observations. Below is a brief summary of the comparison of size, velocity and dust production rates.\\

  \textbf{Main-belt Comets:}
Main-belt comets are objects that show recurrent mass loss near perihelion yet orbit in the main asteroid belt and may be tracers of ice in the inner solar system (see Jewitt \& Hsieh in this volume). Their emission characteristics appear as continuous emission of (sub-)millimeter-sized grains at $<$1 m~s$^{-1}$ speeds (similar to the nuclear escape speed) over weeks to months, with low dust production rates at $<$1 kg~s$^{-1}$.
Similar model parameters were found for several main-belt comets, indicating that these objects share similar properties of the ejected dust \citep{Hsieh2009,Moreno2011,Jewitt2019-S5,Kim2022a,Kim2022b}.
  \\

  \textbf{Jupiter-family Comets:}
Jupiter-family comets typically emit sub-millimeter or larger particles at speeds approximately several tens of m~s$^{-1}$ (slightly lower than the classical \citet{whipple1950} model but still higher than that of main-belt comets) and dust production rates higher than a few hundred kg~s$^{-1}$ near perihelion, although there are variations between individual objects \citep{ishiguro-sarugaku2007, Kelley2008, moreno2009, Agarwal2010}.
  \\

  \textbf{Long-period Comets:}
Long-period comets show a more diverse distribution of dust parameters than 
short-period comets.
The scattering cross-sectional areas of several long-period cometary comae are dominated by micron-sized particles \citep{fulle-cometsII-2004,lisse-ahearn1998}. However, recently observed distant comets show the absence of small particles, and their coma and tails are composed only of particles larger than a millimeter \citep{jewitt19, jewitt_B1_2019}. Dust speed and dust production rates as function of particle size also show large variance.
  \\

  \textbf{Interstellar Comets:}
An accurate determination of the orbit of 1I/`Oumuamua revealed the existence of non-gravitational acceleration, for which the most straightforward explanation would be comet-like outgassing \citep{Micheli2018}. However, the outgassing required to supply the non-gravitational acceleration was predicted to be accompanied by a visible dust or gas coma, while `Oumuamua was always observed as point-like.  The morphology of 2I/Borisov is best reproduced by dust dynamical models if the coma is dominated by sub-millimeter and larger particles, emitted at $\lesssim$9 m~s$^{-1}$ speeds, with total dust production rates  estimated from imaging data $\sim$35~kg~s$^{-1}$ \citep{kim-jewitt2020,Cremonese2020}.

\subsubsection{Neckline}

\begin{figure*}
  \includegraphics[width=\textwidth]{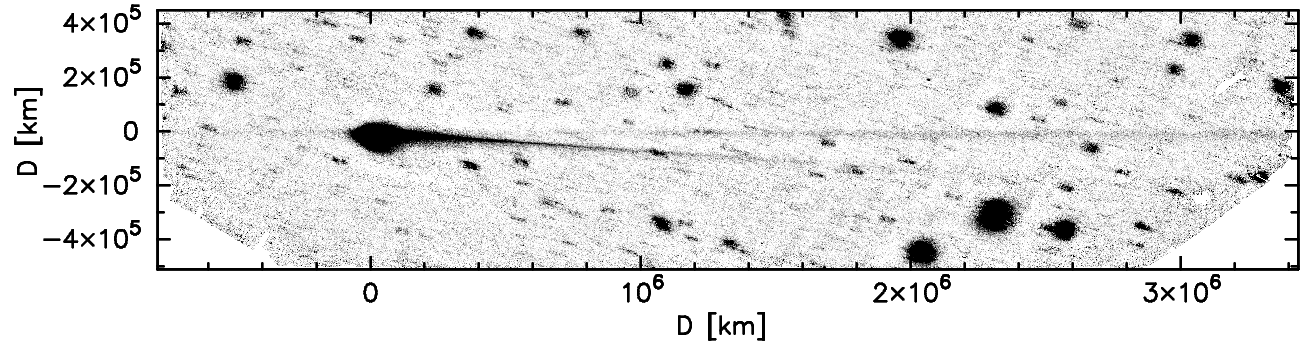}
  \caption{Subaru telescope Hyper Suprime-Cam (HSC) image of comet 67P/Churyumov-Gerasimenko (UT 2016 March 8). The trail (parallel to the horizontal axis) and the neckline structure are projected to different sky position angles. Image courtesy by F. Moreno, original version published as \cite{moreno-munoz2017}.}
  \label{fig:trail}
\end{figure*}

The neckline is a substructure detected in dust tails on rare occasions and caused by dynamical effects \citep{kimura-liu1977}. A neckline consists of large particles emitted at a true anomaly of 180$\degr$ before the observation (cf. Fig.~\ref{fig:trail}) that were emitted with a non-zero velocity component perpendicular to the parent body's orbital plane. Their large sizes imply low $\beta$ and low ejection speeds. Hence their orbits will overall be similar to that of the parent comet, but inclined with respect to it due to the non-zero perpendicular velocity component. After initially dispersing in the perpendicular direction, particles ejected at a given time will re-assemble in the orbital plane of the comet after 180$^\circ$ of true anomaly, and be observable as a thin, bright line of dust.
Using neckline photometry and Monte Carlo models, it is possible to determine if there were significant dust emissions at any given time.
\cite{fulle-barbieri2004} identified comet 67P/Churyumov-Gerasimenko to have a neck-line structure and concluded that the comet has significant dust production at 3.6 au pre-perihelion.
Taking advantage of the neck-line effect, \cite{ishiguro-sarugaku2016} succeeded in detecting the debris cloud ejected from the 2007 outburst of comet 17P/Holmes.

\subsubsection{Trails and dispersion into the zodiacal cloud}
\label{subsec:zodiacal}

Comet debris trails consist of large particles that weakly interact with solar radiation pressure. They were first observed by the Infrared Astronomical Satellite (IRAS) \citep{sykes-lebofsky1986}, and subsequently in both visible and infrared light \citep{ishiguro-watanabe2002, reach07, sykes2004, arendt2014}.
Trails that intersect the Earth's orbit are observed as meteor showers.
As of 2022 January, there are 44 known comets with remotely observed debris trails whose details are available at the website\footnote{ (\url{https://www.astro.umd.edu/~msk/science/trails})} and 24 known comets with dust trails implied by meteor showers (Ye et al., this volume).
Recent models of cometary meteoroid streams show that comet trails can also be observed by in situ dust detectors \citep[e.g.,][]{Kruger2020}.

Comet debris trails contribute a significant input to the interplanetary dust
particle (IDP) cloud complex \citep{sykes2004, reach07}.
\citet{Dikarev2004} and \citet{Soja2019} quantified that the interplanetary dust cloud at 1 au is sustained mainly by Jupiter-family comets ($\sim$90\%), with additional contributions by asteroids ($\sim$10\%), and Halley-type comets ($<$1\%).
\cite{nesvorny-jenniskens2010} connected the vertical brightness profile of the observed mid-infrared zodiacal light with that of a numerical model, suggesting that $\sim$90\% of the zodiacal cloud emission came from comets. \cite{yang-ishiguro2015} reached similar conclusions by comparing the observed optical properties of zodiacal light (i.e., albedo and optical spectral gradients) with those of other types of small bodies in the solar system. In contrast, a non-negligible fraction of asteroid particles in the IDP cloud is proposed by \cite{ipatov-kutyurev2008} and \citet{kawara-matsuoka2017}.

\begin{deluxetable}{llcllcc}

  \tablecaption{Size Distribution Indices\tablenotemark{a}
    \label{size}}
  \tablewidth{0pt}

  \tablehead{ \colhead{Comet} & Method\tablenotemark{b} & Radii ($\mu$m) & Index, $\alpha$ & Reference }
  \startdata

  1P/Halley								& In-Situ & $>$20     & 3.5$\pm$0.2 		  &  \cite{Fulle1995} \\

  2P/Encke								& Optical &  $>$ 1 & 3.2 to 3.6 			    & \cite{Sarugaku2015} \\

  22P/Kopff 								& Optical & $>$ 1  & 3.1				   & \cite{Moreno2012} \\

  26P/Grigg-Skjellerup						& Optical & $>$ 60 & 3.3			            & \cite{Fulle1993} \\

  67P/Churyumov-Gerasimenko (coma)		& In-Situ & $>$ 0.01 & 3.7$_{-0.1}^{+0.7}$  & \cite{Marschall2020}  \\
  67P/Churyumov-Gerasimenko (trail)			& Optical & $>100$   & 4.1			  & \cite{Agarwal2010} \\

  81P/Wild								& Optical & $>$ 1 & 3.45$\pm$0.1 		  & \cite{Pozuelos2014} \\

  103P/Hartley 							& In-Situ & $>$ 10$^4$ & 4.7 to 6.6 		  & \cite{kelley13} \\
  103P/Hartley 							& Optical & $>$ 1 & 3.35$\pm$0.1 		  & \cite{Pozuelos2014} \\

  209P/LINEAR							& Optical & $>$ 1  & 3.25$\pm$0.1              & \cite{Ishiguro2015} \\

  \enddata

  \tablenotetext{a}{Differential power-law size distribution index, $\alpha$. Table adapted from \citet{jewitt-kim2021}.}
  \tablenotetext{b}{In-Situ: direct measurements from spacecraft in the coma.  Optical: determination by fitting tail isophotes in remote sensing data.}

\end{deluxetable}

\section{Future perspectives}
\label{sec:future}

In this chapter, we have attempted to describe our current understanding of how dust is released from a cometary surface and transported to interplanetary space. The vast amount of data returned by spacecraft missions and modern telescopes over the past 20 years, since the publication of the ``Comets II'' book, has shed light on the complexity of this process and left us with a number of open questions that we outline in the following.

One question relates to how activity works at the surface level, that is, how dust and ice are mixed in the cometary nucleus, and which processes lead to the lifting of dust particles, overcoming cohesive forces (Section~\ref{sec:bottleneck}). We need a theoretical description of this process that is not in conflict with the observation that activity exists also and in particular far from the Sun. In our view, current theoretical efforts are limited by the quality of data that we have to constrain them. To characterize the surface composition, texture and structure, highly resolved remote sensing data would be needed, especially at mid-infrared wavelengths, where the maximum of thermal emission in the inner solar system occurs, and in polarized light. In situ analyses of the surface from a landed laboratory would also greatly help to understand the physical and chemical properties of the surface, and finally, experiments with analog materials in an Earth- or space-station based laboratory can provide good constraints as well (see Poch et al. in this book).

The second question addresses the extent to which results obtained from spacecraft missions can be generalized to the wider comet population. On the one hand, this can be addressed by comparing results from the different missions we have had until now, searching for similarities, differences and repeating patterns. Since space missions are costly, the major bridge to the comet population in general will, however, be achieved through telescope observations. To link telescope observations and space missions, we need to understand which properties of the early, near nucleus dust dynamics are still reflected in the outer coma and tail and hence accessible to telescopes. Indications are that much information on the details of the activity distribution is lost in the outer coma \citep{crifo-rodionov1997b, fulle-cometsII-2004}, but remote telescope observations do reveal brightness variations in the outer coma that have not yet been linked to features in the inner coma \citep{knight-snodgrass2017}. A possibility to establish and investigate this connection would be through dedicated modelling of the interface between inner coma and tail for those comets that have been visited by spacecraft. Modelling-wise, these two regions are typically treated separately according to the prevailing forces (gas drag vs. radiation pressure), but see Section~\ref{subsec:sc_and_remote_comae} for examples of connecting spacecraft and telescope observations of cometary comae.

A third complex of open questions concerns how dust evolves in its physical properties while it travels away from the nucleus. Potentially relevant processes include outgassing of embedded ice (affecting both dynamics and physical properties) and fragmentation, possibly induced by outgassing and/or fast rotation, and leading to a change in the dust size distribution. Data do not yet give clear evidence for or against any of these processes. High-resolution ground-based observations using complementary techniques such as visible light and thermal infrared spectroscopy and polarimetry could provide stronger constraints on the dust evolution at least in the outer coma.


\vskip .5in
\noindent \textbf{Acknowledgements.} \\

We thank Vladimir Zakharov, David Jewitt, Xian Shi, Felix Keiser, Toshi Kasuga, Marius Pfeifer, and the referees, Eberhard Gr\"un and Jean-Baptiste Vincent, for their comments that significantly helped us to improve this manuscript.
J.A. and Y.K. acknowledge funding by the Volkswagen Foundation. J.A.'s contribution was made in the framework of project CAstRA funded by the European Union's Horizon 2020 research and innovation program under grant agreement No. 757390.
M.S.P.K. acknowledges support from NASA Grant 80NSSC20K0673.

\bibliographystyle{sss-three.bst}

\bibliography{references}

\end{document}